\documentclass[10pt,a4paper,twoside]{article}
\usepackage[utf8]{inputenc}
\usepackage[T1]{fontenc}
\usepackage[english]{babel}

\usepackage{fancyhdr}
\usepackage[usenames,dvipsnames,svgnames,table]{xcolor}
\definecolor{citegreen}{rgb}{0.00,0.70,0.30}
\usepackage[colorlinks=true,
            linkcolor=blue,
            urlcolor=blue,
            citecolor=blue]{hyperref}

\usepackage{amsrefs}
\usepackage{amssymb}
\usepackage{amsmath}
\usepackage{amsthm}
\usepackage{mathrsfs}

\usepackage{xspace}

\usepackage{pgf}
\usepackage{tikz}

\usetikzlibrary{arrows,automata}

\usetikzlibrary{calc}

\DeclareMathAlphabet{\mathpzc}{OT1}{pzc}{m}{it}

\usepackage{epsfig}
\usepackage{enumerate}
\usepackage{graphicx}

\usepackage[title]{appendix}

\makeatletter 
\renewcommand\@biblabel[1]{#1.} 
\makeatother

\numberwithin{equation}{section}

\newtheorem{theorem}{Theorem}[section]

\newtheorem{lemma}[theorem]{Lemma}
\newtheorem{corollary}[theorem]{Corollary}
\newtheorem{remark}{Remark}[section]

\newtheorem{definition}{Definition}[section]

\definecolor{redd}{rgb}{0.95,0.2,0.2}
\definecolor{gris}{rgb}{0.9,0.9,0.9}
\definecolor{greenn}{rgb}{0.1,0.6,0.2}

\definecolor{cmgray}{rgb}{0.7,0.7,0.7}

\definecolor{cmblue}{rgb}{0.2,0.5,0.8}

\providecommand{\B}{\mathbf}

\providecommand{\D}{\mathbb}

\providecommand{\R}{\mathrm}

\newcommand{\eu}{\mathrm{e}}

\def\BI{{\mathbf I}}
\def\bf1{{\mathbf 1}}

\def\bcB{\boldsymbol{\mathcal{B}}}

\def\bcBN{\mathbf{B}^{(N)}}

\def\dist{\mathrm{dist}}

\def\one{\mathbf{1}}

\def\mes{{\rm mes}}

\def\diam{{\rm diam}}

\def\rc{{\complement}}

\def\lam{{\lambda}}

\def\eps{\epsilon}

\def\Bigesm#1{\D{E}\Big[\, #1\, \Big]}

\def\lr#1{\langle#1\rangle}

\def\SI{{\rm SI}\xspace}

\def\SSI#1{\textsf{S}$(#1)$}

\def\SSIexp#1{$\textsf{S}_{\text{\sc{exp}}}(#1)$}

\def\EbR{$E$-R\xspace}

\def\FNR{$E$-FNR\xspace}

\def\EmS{$(E,m_N)$-S\xspace}
\def\EmNS{$(E,m_N)$-NS\xspace}

\def\EdS{{\rm$(E,\delta)$-S}\xspace}

\def\EdmNS{$(E,\delta,m)$-{\rm{NS}}\xspace}
\def\EdmS{$(E,\delta,m)$-{\rm S}\xspace}

\def\mdmS#1{$(#1)$-S\xspace}
\def\mEdmS#1{$(E,#1)$-{\rm S}\xspace}
\def\mEdmNS#1{$(E,#1)$-{\rm{NS}}\xspace}

\def\EbR{$(E,\beta )$-{\rm R}\xspace}
\def\EbNR{$(E,\beta )$-{\rm{NR}}\xspace}

\def\htil{{\widetilde{h}}}

\def\brho{{\boldsymbol{\rho}}}
\def\brhoS{{\boldsymbol{\rho}_{\R{S}}}}

\def\hBx{\widehat{\Bx}}

\providecommand{\bS}[1]{\boldsymbol{#1}}

\def\BA{\mathbf{A}}

\def\rBF{{\mathrm{\mathbf{F}}}}
\def\BG{\mathbf{G}}
\def\BH{\mathbf{H}}

\def\BHBni{\mathbf{H}^{\rm ni}_{\bcB}}

\def\BGBni{\mathbf{G}_{\bcB}^{\rm ni}}

\def\BU{\mathbf{U}}

\def\BUJ{\BU_{\cJ, \cJ^\rc}}

\def\BA{\mathbf{A}}

\def\Bx{\mathbf{x}}
\def\By{\mathbf{y}}
\def\Bu{\mathbf{u}}
\def\Bv{\mathbf{v}}
\def\BW{\mathbf{W}}

\def\Bz{\mathbf{z}}

\def\Bpsi{{\boldsymbol{\psi}}}
\def\Bphi{{\boldsymbol{\phi}}}
\def\PPi{{\mathrm{\Pi}}}

\def\BPsi{{\bS{\Psi}}}

\def\DC{\D{C}}

\def\DR{\D{R}}
\def\DZ{\mathbb{Z}}
\def\DN{\D{N}}

\def\cB{\mathcal{B}}

\def\csB{\mathscr{B}}

\def\csR{\mathscr{R}}
\def\csS{\mathscr{S}}

\def\cC{\mathcal{C}}
\def\cD{\mathcal{D}}
\def\cE{\mathcal{E}}

\def\cZ{\mathcal{G}}

\def\cJ{\mathcal{J}}
\def\cL{\mathcal{L}}

\def\csP{\mathscr{P}}
\def\csQ{\mathscr{Q}}
\def\cQ{\mathcal{Q}}

\def\cS{\mathcal{S}}

\def\cV{\mathcal{V}}

\def\cZ{\mathcal{Z}}

\def\rA{{\R{A}}}

\def\rBF{{\mathbf{F}}}

\def\rM{{\R{M}}}

\def\rP{{\R{P}}}

\def\rd{{\R{d}}}

\def\rW{{\R{W}}}

\def\rP{{\R{P}}}
\def\rQ{{\R{Q}}}
\def\rR{{\R{R}}}
\def\rS{{\R{S}}}

\def\rw{{\R{w}}}

\def\hN{{N^*}}

\def\mnu{\nu}

\def\be{\begin{equation}}
\def\ee{\end{equation}}
\def\ba{\begin{array}{l}}
\def\ea{\end{array}}
\def\bal{\begin{aligned}}
\def\eal{\end{aligned}}

\def\ble{\begin{lemma}}
\def\ele{\end{lemma}}

\def\bthm{\begin{theorem}}
\def\ethm{\end{theorem}}

\def\bco{\begin{corollary}}
\def\eco{\end{corollary}}

\def\bre{\begin{remark}}
\def\ere{\end{remark}}

\def\btm{\begin{theorem}}
\def\etm{\end{theorem}}

\def\bde{\begin{definition}}
\def\ede{\end{definition}}

\def\fF{\mathfrak{F}}
\def\fG{\mathfrak{G}}

\def\fS{\mathfrak{S}}
\def\fs{\mathfrak{s}}

\def\om{{\omega}}
\def\Om{{\Omega}}

\def\eps{\epsilon}

\def\lam{{\lambda}}

\def\bcB{{\boldsymbol{\mathcal{B}}}}

\def\bcE{{\boldsymbol{\mathcal{E}}}}

\def\bcZ{{\boldsymbol{\mathcal{G}}}}
\def\bcZN{{\boldsymbol{\mathcal{G}^{N}}}}

\def\bcL{{\boldsymbol{\mathcal{L}}}}

\def\bcN{{\boldsymbol{\mathcal{N}}}}

\def\bcS{{\boldsymbol{\mathcal{S}}}}

\def\bcU{{\boldsymbol{\mathcal{U}}}}
\def\bcV{{\boldsymbol{\mathcal{V}}}}
\def\bcW{{\boldsymbol{\mathcal{W}}}}

\def\bcZ{{\boldsymbol{\mathcal{Z}}}}
\def\bcZN{{\boldsymbol{\mathcal{Z}^{N}}}}

\def\pr#1{\D{P}\left\{\,#1\,\right\}}
\def\esm#1{\D{E}\left[\, #1\, \right]}
\def\pt{\partial}
\def\half{\frac{1}{2}}

\def\quart{\frac{1}{4}}

\def\nb#1{{ \langle #1 \rangle}}

\def\truc#1#2#3{\smash{\mathop{\,\, #1 \,\, }\limits^{#2}_{#3}}}

\def\diy{\displaystyle}

\def\Uone{\textsf{(U)}\xspace}

\def\Vone{\textsf{(V)}\xspace}

\def\RCM{\textsf{(RCM)}\xspace}

\begin{document}

\title{Efficient Anderson localization bounds \\ for large multi-particle systems}


\author
{Victor Chulaevsky, Yuri Suhov}


\date{}

\maketitle
\begin{abstract}
We study multi-particle interactive quantum disordered systems on a
polynomially-growing countable connected graph $(\cZ,\cE)$. The novelty is to give
localization bounds uniform in finite or infinite volumes (subgraphs) in $\cZ^N$ as well
as for the whole of $\cZ^N$. Such bounds are proved here by means of a
comprehensive  fixed-energy multi-particle multi-scale analysis.
Another feature of the paper is that we consider -- for the first time in the
literature -- an infinite-range (although fast-decaying) interaction between
particles.  For the models under consideration we establish (1) exponential
spectral localization, and (2) strong dynamical localization with sub-exponential
rate of decay of the eigenfunction correlators.
\end{abstract}

\thispagestyle{empty}

\section{Introduction. The model and results}  
\label{Sec:intro}

Until recently, the rigorous Anderson localization theory focused on single-particle models.
(In the physical community,  notable papers on multi-particle systems with interaction
appeared as early as in 2005--2006; see \cite{GorMP05,BAA06}.)

Initial rigorous results on multi-particle lattice localization for a finite-range two-body interaction
potential were presented in \cite{CS08,CS09a,CS09b} and \cite{AW09a,AW09b};
continuous models have been considered in \cite{BCSS10}, \cite{CBS11}, and later in \cite{HK13}, \cite{Sab13}.
In these papers, both Spectral localization (SL) and Dynamical Localization (DL) have
been established. A considerable progress was made in \cite{KN13a,KN13b}, with the help of
an adapted bootstrap variant of the Multi-Scale Analysis (MSA) developed
in the earlier work  \cite{GK01}. The resulting bootstrap multi-particle MSA (MPMSA)
was applied
in [24] and [25] to multi-particle systems in the lattice and in the Euclidean
space, respectively. More recently, a very important step was made in the paper
\cite{FW14} which extended the multi-particle
Fractional-Moment Method (MPFMM) from the lattice case \cite{AW09a,AW09b}  to the continuous
one, with an infinite-range two-body interaction potential. As usual, (MP)FMM
provides, under certain assumptions, exponential decay bounds upon the eigenfunction
correlator (EFC), while the bootstrap MPMSA achieves only a sub-exponential
decay of the EFC at large distances.

The main motivation for the present work comes from the fact that in all
above-mentioned papers the decay bounds on the eigenfunctions and
EFCs was proved in the so-called Hausdorff distance (HD) which is actually,
a pseudo-distance in the multi-particle configuration space. In the context of the
multi-particle Anderson localization, the HD appears explicitly
in \cite{AW09a,AW09b} (as well as in \cite{KN13a,KN13b}), while in \cite{CS08,CS09a,CS09b}
it was used implicitly, through
the notion of separated cubes. The point is
that there are arbitrarily distant loci in the multi-particle space
which might support quantum tunneling between them, and the HD does not
reflect this possibility. Another point is that the SL and DL
have been proved so far in an
infinitely extended physical configuration space, but some tunneling processes
could not be ruled out in arbitrarily large, yet bounded subsets thereof. As a result,
the existence of efficient multi-particle localization -- even for a bounded number
of particles $N\ge 3$ -- remained an open question.
These aspects of the rigorous multi-particle localization theory were analyzed by Aizenman and
Warzel in \cite{AW09a,AW09b}, and their analysis of the problem was instrumental for a
partial solution given in  \cite{C12b,C10a}. The mathematical core of the problem
is an eigenvalue concentration (EVC) bound for two distant loci in the multi-particle
space, used in different ways in the MPMSA and in MPFMM. In the current paper we
employ a probabilistic result from \cite{C13a} and prove a suitable EVC bound (cf. Theorem
2.2) for a class of sufficiently regular marginal probability distributions of IID
external random potentials. We expect such a bound to be extended to
a larger family of random potentials.
It has to be emphasized that the problem in question appears only for the
number of particles $N\geq 3$, and the proof of localization for two-particle systems
given in \cite{CS09a} operates with the (symmetrized) norm-distance in
the two-particle space. As a result, the two-particle localization holds
in finite (but arbitrarily large) regions of the physical configuration space, under
mild regularity conditions upon the random potential; see \cite{CS13}.

In the present paper we focus on an interactive $N$-particle Anderson
model, on a countable connected graph $(\cZ;\cE)$ with a polynomially growing
size of a ball when the radius increases to infinity.  The main method used is
a new variant of the MPMSA. The results are summarized as follows.
\begin{itemize}
\item We prove uniform localization bounds, in terms of decay of eigenfunctions
(EFs) and eigenfunction correlators (EFCs) valid for
finite or infinite subgraphs of $\cZ^N$, including  the whole $\cZ^N$.
Previously published results provided less efficient bounds in finite
volumes.
\item As in \cite{FW14}, we treat systems with infinite-range interaction
potentials. Specifically, we
consider a two-body potential decaying at a large distance $r$ as $\eu^{-r^\zeta}$ where
$\zeta > 0$. Surprisingly, the SL holds here with an exponential
rate ($\eu^{-mr}$, $m > 0$) even if $0<\zeta < 1$.
We want to note that an exponential decay of EFs was proved in \cite{C12a}
under the assumption of
decay of the interaction with rate $\eu^{-r^\zeta}$, but only for $\zeta\in (0; 1]$ sufficiently close
or equal to $1$. Paper \cite{FW14} established the EFC decay with rate $\eu^{-\kappa r}$, $\kappa\in (0; 1]$, in the
following three cases:
   \begin{itemize}
   \item The interaction potential decays at an exponential rate $\eu^{-ar}$; in this case the EFCs also decay exponentially
    fast.
    \item The interaction potential decays sub-exponentially, as $\eu^{-r^\zeta}$ with $\zeta\in (0,1)$;
     in this case the EFCs also decay sub-exponentially.
    \item The interaction potential decays polynomially, as $C r^{-A}$, with a sufficiently large $A > 0$); in this case the
     EFCs also decay sub-exponentially, i.e., at a much faster rate than the interaction.
   \end{itemize}
\end{itemize}

We present competing bounds for the EFCs. As was said, we
also show \emph{exponential} decay of EFs. Together with the results of
\cite{C12a} and \cite{FW14}, this evidences that the decay rate of the EFs and EFCs
can be stronger than that of the interaction potential.
The EFC decay bounds are established here
in the natural (symmetrized) norm-distance, more efficient than
the Hausdorff pseudo-distance.

The rest of the paper is devoted to the proof of Theorem 1.1. In particular,
in Section 2 we establish an important ingredient of the proof: eigenvalue
concentration bounds (EVCs). The bulk of the work is about the proof of assertion
(A): it is carried in Section 3. The main strategy here is the induction on the
number of particles $N$, initially developed in \cite{CS09a,CS09b}. Each step $N\rightsquigarrow  N + 1$,
$N = 1,\ldots ,N^* -1$, employs the multi-scale analysis of multi-particle Hamiltonians.
Unlike Ref. \cite{CS09b}, we make use of a more efficient scaling technique, essentially
going back to the work \cite{GK01} and recently adapted in \cite{KN13a} to multi-particle systems.

We need to modify the bootstrap MPMSA strategy from \cite{KN13a}. Specifically, we carry out
only two of the four separate scaling analyses which constitute the bootstrap
method. This results in a shorter proof of sub-exponential decay of the EFCs.
The full-fledged bootstrap MPMSA (cf. \cite{KN13a}),
combined with our new eigenvalue concentration estimate (cf. Theorem 2.2),
allows us to prove the EFC decay in the symmetrized graph distance.
It is worth mentioning that the base of induction ($N = 1$) requires a proof of
localization bounds for single-particle systems on graphs with a polynomial growth
of the size of a ball with the radius. The required estimates were proved in
Ref. \cite{C14} following the techniques from \cite{GK01}.

\def\ssect{\subsection}

\ssect{The multi-particle Hamiltonian}\label{ssect:LaplHsam} 

Consider a finite or locally finite, connected non-oriented graph $(\cZ, \cE)$,
with the vertex set $\cZ$ and the edge set $\cE$. (For brevity, we often refer to $\cZ$ only.)
We assume that $\cE$ does not include
cyclic edges $x\leftrightarrow x$ and denote by $\rd (\,\cdot\,,\,\cdot\,)$
the graph distance on $\cZ$: $\rd (x,y)$ equals the length of the shortest path
$x\leftrightsquigarrow y$ over the edges. (By definition, $\rd (x,x)=0$.)
We assume that graph $(\cZ,\cE)$ belongs to a class $\fG (d,C)$ for some $d,C>0$, meaning
that the size of a ball $\cB(x,L):=\{y:\, \rd (x,y)\le L\}$ is polynomially bounded:
\be\label{eq:ball.growth}
\sup_{x\in\cZ}\sharp \cB(x,L) \le C L^d, \;\; L\ge 1.
\ee
Physically, $\cZ$ represents  the configuration space
of a single quantum particle.
\medskip

\textbf{NB.} To keep the track of the constants emerging in the course of the presentation,
we will use a sub-script indicating their origin; e.g., $C_\cZ$  will refer to the constant(s)
arising in connection with graph $(\cZ,\cE)$ as in the paragraph above. $\Box$
\medskip

The configuration space of $N$ distinguishable particles is the graph $(\cZ^N,\cE_N)$.
Here $\cZ^N$ is the Cartesian power, and  the edge set $\cE_N$ is defined as follows.
Given $\Bx =(x_1,\ldots ,x_N),\By =(y_1,\ldots ,y_N)\in\cZ^N$, the edge $\Bx\leftrightarrow\By$
exists if, for some $j=1,\ldots ,N$, there exists an edge  $x_j\leftrightarrow y_j$ in $\cE$
while for $i\neq j$ we have $x_i=y_i$. We refer to $\Bx$ and $\By$ as $N$-particle
configurations (briefly, configurations) on $\cZ$ and use the same notation
$\rd (\Bx,\By )$ for the graph distance on $\cZ^N$ (as before, $\rd(\Bx,\Bx )=0$).
Apart from the distance $\rd(\cdot\,,\,\cdot)$
on $\cZ^N$, it will be convenient to use the max-distance $\brho$ and the symmetrized max-distance $\brhoS$, defined
as follows:
\be\label{eq:metrics}
\brho(\Bx, \By) = \max_{1\le j \le N} \rd (x_j, y_j);
\quad
\brhoS(\Bx, \By) = \min_{\pi\in\fS_N} \brho\big(\Bx, \pi(\By) \big).
\ee
Here the symmetric group $\fS_N$ acts on $\cZ^N$ by permutations
of the coordinates.

Next,
$\bcB^{(N)}(\Bx,L)$ denotes the ball in $\cZ^N$ centered at $\Bx=(x_1,\ldots ,x_N)$ in metric $\brho$
(sometimes called an $N$-particle ball):
\be\label{eq:ball}
\bcB^{(N)}(\Bx,L):=\{\By:\, \brho (\Bx,\By)\le L\}=\operatornamewithlimits{\times}_{j=1}^N \cB (x_j,L).\ee
It will be often convenient to omit the index $N$ and use the boldface notation:
$\bcZ=\cZ^N$, $\bcE=\cE_N$, $\bcB(\Bx,L)=\bcB^{(N)}(\Bx,L)$, etc.
Note that for $\cZ\in\fG(d,C)$, one has
$\sharp\,\pt \bcB^{(N)}(\Bx ,L) \le C^{2N} L^{Nd}$.

The graph Laplacian $\Delta_\cZ$ on $\cZ$ is given by
\be\label{eq:def.Laplace.graph}
(\Delta_\cZ f)(x) = \sum_{ \nb{x,y} } (f(y) - f(x))
= - n_\cZ(x) f(x) + \sum_{ \nb{x,y} } f(y),\;x\in\cZ;
\ee
here $\nb{x,y}$ stands for a pair $(x,y)\in\cZ\times\cZ$  with $\rd (x,y)=1$, and $n_\cZ(x)$\\ $=\sharp\,\{y:
\rd (x,y)=1\}$. Similarly to
\eqref{eq:def.Laplace.graph}, the Laplacian on $\bcZ$ is defined by
\be\label{eq:def.Laplace.graphN}
\begin{array}{cl}
(\Delta_\bcZ f)(\Bx)&=\diy\sum\limits_{1\leq j\leq N}(\Delta^{(j)}_\cZ f)(\Bx) \\
\;&= \diy\sum\limits_{ \nb{\Bx,\By} } (f(\By) - f(\Bx))
= - n_\bcZ(\Bx) f(\Bx) +\diy \sum\limits_{ \nb{\Bx,\By} } f(\By),\Bx\in\bcZ.
\end{array}\ee
Here $\Delta^{(j)}_\cZ$ denotes the Laplacian acting on the $j$th component of $\Bx$,
$\nb{\Bx,\By}$ stands for a pair $(\Bx,\By)\in\bcZ\times\bcZ$  with $\rd (\Bx,\By)=1$ and
$n_\bcZ(\Bx)=\sharp\;\{\By:\;\rd (\Bx,\By)=1\}$.

Given $\cV\subset\cZ$, we write $\nb{x,y}\in\cV$ meaning that
$x,y\in\cV$ and $\rd(x,y)=1$. Likewise, for
$\bcV\subseteq\bcZ$ the notation $\nb{\Bx,\By}\in\bcV$ means
$\Bx,\By\in\bcV$ and $\rd(\Bx,\By )=1$. With this agreement, the Laplacians
$\Delta_\cV$ and $\Delta_\bcV$ (with Dirichlets boundary condition)
are introduced as follows:
\be\label{eq:def.Laplace.graphB}
\Delta_\cV = \one_{\cV} \Delta_\cZ \one_{\cV}
\ee
and
\be\label{eq:def.Laplace.graphNB}
\Delta_\bcV = \one_{\bcZ} \Delta_\bcZ \one_{\bcZ}.
\ee

The $N$-particle Hamiltonian
$\BH^{(N)}_\bcV=\BH^{(N)}_\bcV(\omega )$ in `volume` $\bcV\subseteq\bcZ$ acts as
\begin{equation}\label{HamVN}\begin{array}{cl}
\left(\BH^{(N)}_\bcV f\right)(\Bx )&=\left(-\Delta_\bcV f\right)(\Bx)+g\diy\sum\limits_{1\leq j\leq N}
V(x_j;\omega )f(\Bx)\\
\;&+\diy\sum\limits_{1\leq i<j\leq N}U(\rd (x_i,x_j))f(\Bx ),\;\Bx =(x_1,\ldots ,x_N)\in\bcV.\end{array}
\end{equation}
Here $V$ represents a random external field and $U$ a two-body interaction; see below. The constant
$g\in\DR$ is referred to as a coupling amplitude.
Under the imposed conditions,  with probability one, the operator $\BH^{(N)}_\bcV(\om)$ is bounded and self-adjoint
in $\ell_2(\bcV)$.

\ssect{The assumptions and results}  
\label{ssec:assump.V}

Our goal is to prove that, under certain conditions on $V(\cdot;\om)$ (the external potential) and  $U(\cdot)$
(the two-body interaction potential)
and for sufficiently large values of the disorder amplitude $|g|$,
the (random) eigenvectors  of $\BH^{(N)}_\bcV$ in $\ell_2(\bcV)$
feature strong decay properties, stated in appropriate terms. We stress
that we establish the threshold  for $|g|$ (and other bounds involved) uniformly in
$\bcV$ for a bounded range of values of $N$.  Formal statements are given in Theorem \ref{thm:Main.DL.Vone.Uone.strong.g} below. \def\unp{{\underline p}}  \def\ovp{{\overline p}}

The condition upon $V$ is:

\vskip2mm
\noindent
\Vone:
\textit{The random field
$(x,\omega )\mapsto V(x;\omega)\in\DR$ 
is {\rm IID}, with a marginal probability distribution
supported by a bounded interval and admitting a smooth probability density $p_V$
satisfying the following conditions: $\;\forall\;t\in{\rm{supp}}\;p_V$,}
\be\label{eq:condsV}
0 <\unp \le p_V(t) \le \ovp < \infty;\quad
|p^\prime_V(t)|\leq R<\infty .\ee

The probability distribution generated by the random variables $V(x)$ is
denoted by $\mathbb P$ and the expectation by $\mathbb E$.

We assume the following condition upon $U$.
\vskip2mm

\noindent
\Uone: \ \textit{$\;\exists$ $\zeta >0$ and $C=C_U >0$ such that
\be\label{eq:def.U1}
|U (r) | \le C\eu^{ - r^{\zeta}},\;\;r=1,2,\ldots .
\ee
}

When $\zeta\in (0,1)$, it makes sense to refer to a sub-exponential decay of $U$,
and with $\zeta=1$ we have the case of an exponential decay.

Let $\csB_1$ denote the set of all continuous functions $f: \DR\to\DC$ with $\|f\|_\infty\le 1$.
The results of this paper are presented in Theorem \ref{thm:Main.DL.Vone.Uone.strong.g}.

\begin{theorem}\label{thm:Main.DL.Vone.Uone.strong.g}
Assume Conditions {\rm\Vone} and {\rm\Uone} and fix an integer ${\hN}\ge 2$.
$\exists$ $\kappa^*=\kappa^*(\zeta,N^*) \in (0,\zeta ]$ with the following property.
$\forall$  $\kappa\in(0, \kappa^*]$ and $m>0$,
there is a value $g_0=g_0(\hN,m,\kappa)>0$ such that $\forall$ $N=1,\ldots ,{\hN}$
and $|g|\in (g_0,+\infty )$:
\par\noindent
{\rm{\textbf{(A)}}}
$\exists$ a constant $C=C_{\rm EFC}>0$ such that $\forall$ set $\bcV\subseteq\bcZ$
(finite or infinite) and $\forall$ $\Bx,\By\in\bcV$,
\be\label{eq:thm.Main.DL.02}
\Bigesm{ \sup_{f\in\csB_1} \big| \langle \one_{\By}\, | \,  f(\BH^{(N)}_\bcV) \, | \,
\one_{\Bx}\rangle\big| }
\le C\, \eu^{ -m\left(\brho_{\rS}(\Bx,\By)\right)^\kappa }.
\ee
\noindent
{\rm{\textbf{(B)}}}
With probability one, $\BH^{(N)}_\bcZ(\om)$ has pure point spectrum.
and all its eigenfunctions
$\BPsi_j(x;\om)$ decay exponentially fast: there exists a nonrandom
number $m=m_N>0$  such that
$\forall$ $\BPsi_j$ $\;\exists$ a constant $\;C_j=C_j(\om )$ and a site
$\hBx_j=\hBx_j(\om)\in\bcZ$ (a localization center) such that
\be\label{eq:thm.Main.DL.01}
| \BPsi_j(\Bx,\om) | \le C_j(\om) \,\eu^{ -m \brho_{\rS}(\Bx,\hBx_j) },\;\;\Bx\in\bcV.
\ee
\vskip 1 truemm\par\noindent
\end{theorem}
\def\rS{{\rm S}}

The quantity in the LHS of \eqref{eq:thm.Main.DL.02} is called the EF correlator (EFC), between $\Bx$ and $\By$,
for Hamiltonian $\BH^{(N)}_{\bcV}$.
Compared with \cite{AW09a,CS09b,KN13a}, Eqns \eqref{eq:thm.Main.DL.01}--\eqref{eq:thm.Main.DL.02}
show the decay in a more suitable form involving metric $\brho_\rS$ rather than the Hausdorff distance.

The rest of the paper is devoted to the proof of Theorem \ref{thm:Main.DL.Vone.Uone.strong.g}.
In particular, in Section \ref{Sec:EVC} we establish an important ingredient of the proof: eigenvalue concentration bounds.
The bulk of the work is about the proof of assertion {\textbf{(A)}}: it is carried in Section \ref{Sec:FEMSA.subexp}.
The main strategy here is the induction on the number of particles $N$, initially developed in
\cite{CS09a,CS09b}. Each step $N \rightsquigarrow N+1$,  $N=1, \ldots, N^*-1$, employs the multi-scale analysis
of multi-particle Hamiltonians. Unlike Ref. \cite{CS09b}, we make use of a more efficient
scaling technique, essentially going back to the work by Germinet and Klein \cite{GK01} and recently
adapted by Klein and Nguyen \cite{KN13a} to multi-particle systems.
However, we do not follow closely the bootstrap (MP)MSA strategy from \cite{KN13a}. Specifically,
we carry out only two of the four separate scaling analyses which constitute the
bootstrap method. This results in a shorter proof of sub-exponential decay of the EFC, with rate
$\eu^{-L^\kappa}$ for some $\kappa>0$. The full-fledged bootstrap MPMSA (cf. \cite{KN13a}), combined with our new eigenvalue
concentration estimate  (cf. Theorem \ref{thm:W2.Vone}),
would allow one to prove the EFC decay in the symmetrized graph-distance (and not only in the Hausdorff distance),
with rate $\eu^{-L^\kappa}$ for $\kappa\in(0,1)$ arbitrarily close to $1$,
while starting from fairly weak assumptions upon the localization properties in the balls of radius
$L_0$.

It is worth mentioning that the base of induction ($N=1$) requires a proof of localization bounds
for single-particle systems on graphs of polynomial growth of balls, and we cannot simply refer
to \cite{KN13a} where, formally speaking, only the lattice systems (on $\DZ^d$, $d\ge 1$) were studied.
The required estimates for the $1$-particle Anderson models on graphs were proved in Ref. \cite{C14}, where it
was emphasized that the main scaling technique is due to Germinet and Klein \cite{GK01}.

The proof of assertion {\textbf{(B)}} is contained in Section \ref{Sec:exp.decay.EF} (it makes use of a number of facts established in other sections). This
proof is based on the (modified) version of the MPMSA presented in \cite{CS13}.
In particular, the case of the two-body potential $U$ satisfying \Uone is treated as a small
perturbation of a finite range interaction. Some technical proofs are presented in the Appendix.
Others repeat arguments published elsewhere (sometimes with minor changes) and are
omitted.

From here on we fix a positive integer $N^* \ge 2$ and consider $N=1,\ldots ,N^*$ without
stressing it every time again.  The
dependence of various quantities upon the upper-bound value $N^*$ is not  emphasized
but of course is crucial throughout the whole construction. Conditions \Vone and \Uone
are also globally assumed,  although a number of intermediate assertions (particularly in
Section \ref{Sec:EVC}) require more liberal restrictions upon $V$.


\section{Eigenvalue concentration bounds}  
\label{Sec:EVC}

\ssect{The resolvent inequalities. Singular and resonant sets} 
\def\BI{{\mathbf I}}

The main tool in the proof of Theorem \ref{thm:Main.DL.Vone.Uone.strong.g} are
properties of decay of the Green functions (GFs) of Hamiltonian $\BH^{(N)}_\bcV$:
$$(\Bx,\By)\in\bcV\times\bcV\mapsto G_\bcV(\Bx, \By)\left(=G^{(N)}_\bcV(\Bx, \By; E; \om)\right),$$
for a finite set $\bcV\subset\bcZ$ and the value $E\not\in\Sigma\left(\BH^{(N)}_\bcV\right)$.
As usual, $G_\bcV(\Bx, \By)$ denotes the matrix entry
of the resolvent $\BG_\bcV(E)=\BG^{(N)}_\bcV(E; \om)$ in the delta-basis :
$$G_\bcV(\Bx, \By)=\left\langle\bf1_\Bx,\left(\BH^{(N)}_\bcV-E\BI\right)^{-1}\bf1_\By\right\rangle\,.$$
The base for the argument is the Geometric resolvent inequality (GRI)
for the GFs: $\forall$ subset $\bcW\subset\bcV$ and configurations $\Bx\in\bcW$,
$\By\in\bcV\setminus\bcW$,
\be\label{eq:GRE}
|G_\bcV(\Bx,\By)| \leq \sum_{ \nb{\Bu,\Bv}\in\pt_\bcV\bcW}\left|G_\bcW(\Bx,\Bu)\right|\,
\left|G_{\bcV}(\Bv,\By)\right|
\ee
Here $\pt_\bcV\bcW$ stands for the edge-boundary of $\;\bcW\;$ in $\;\bcV$:
$$
\pt_\bcV\bcW =\left\{(\Bu,\Bv):\;\Bu\in\bcW,\;\Bv\in\bcV\setminus\bcW,\;
\brho (\Bu,\Bv)=1\right\}.
$$

The distance $\dist$ below refers to the standard metric on the line $\DR$.
The inner boundary $\pt^-\bcV$ is determined by
$$\pt^-\bcV =\{\Bu\in\bcV:\;\brho (\Bu,\bcZ\setminus\bcV)=1\}.$$

\begin{definition}\label{def:NR.NS}
Given $E\in\DR$, $\beta, \delta\in(0,1)$ and $m>0$
an $N$-particle ball $\bcB =\bcB^{(N)}(\Bx ,L)\subset\bcZ$ is called
\begin{itemize}
  \item $(E,\beta )$-resonant (\EbR, in short), if
  \be\label{eq:def.NR}\dist\left(\Sigma\left(\BH^{(N)}_{\bcB}\right), E\right) < 2\eu^{-L^\beta}\ee
  and $(E,\beta )$-non\-resonant (\EbNR), otherwise;
  \item $(E,\delta,m)$-nonsingular (\EdmNS), if for all configurations
  $\By\in\pt^- \bcB$
\be\label{eq:def.NS}
\left|G^{(N)}_{\bcB}(\Bx,\By;E)\right| \le \left({C_\cZ}^{2N} L^{Nd}\right)^{-1}\eu^{-m L^\delta},
\ee
and $(E,\delta,m)$-singular (\EdmS), otherwise.
\end{itemize}
\end{definition}
\def\rB{{\rm B}}

Typically,  properties $(E,\delta,m)$-NS and $(E,\delta,m)$-S
will be used with $m=m_N$ where $m_N$ varies in a certain specified manner
(see \eqref{eq:table1} and \eqref{eq:table2}).

\ssect{One- and two-volume  EVC bounds}

We start with a one-volume EVC bound that is an analog of the
well-known Wegner-type estimate:

\begin{theorem}
\label{thm:W1} Fix $\beta \in (0,1)$
There exists a constant $C=C^{(1)}_{\beta,V}$ such that
$\forall$ $E\in\DR$, $1\leq N\leq N^*$, $\Bx\in\bcZ$ and integer $L>1$,
\be\label{eq:thm.W1}
\sup_{E\in\DR}  \;
\pr{ \text{{\rm{ball}} $\bcB^{(N)}(\Bx ,L)$ {\rm{is}} {\rm \EbR} } } \le C\eu^{-L^{\beta/2} }.
\ee
\end{theorem}
\def\ov{\overline}

The proof of Theorem \ref{thm:W1} is omitted: it repeats the one given in \cite{CS13}, Theorem 3.4.1 (Eqn (3.41)).
Theorem \ref{thm:W1}
is used in the proof of Theorems \ref{thm:MSA.main.fixed} and \ref{thm:MSA.exp.fixed}.

A (new) two-volume EVC is the subject of Theorem \ref{thm:W2.Vone} below. It is instructive to
compare it with Theorem 3.4.2 (Eqn (3.44)),
Corollary 3.1 (Eqns (3.47)--(3.48)) and Theorem 3.5.2 (Eqn (3.58)) in \cite{CS13}.

Given an integer $R\ge 0$, we will say that two balls $\bcB^{(N)}(\Bx,L)$, $\bcB^{(N)}(\By,L)$ are $R$-distant
if
\be
\brhoS(\Bx, \By) \ge R.
\ee

\begin{theorem}\label{thm:W2.Vone}
There is a constant $C=C^{(2)}_V$ such that
for all $s>0$, integer $L>1$ and any pair of $3NL$-distant balls $\bcB^{(N)}(\Bx,L)$,
$\bcB^{(N)}(\By,L)$, the  spectra
$\Sigma_\Bx:=\Sigma(\BH^{(N)}_{\bcB (\Bx,L)})$, $\Sigma_\By:=\Sigma(\BH^{(N)}_{\bcB (\By,L)})$
obey
\be\label{eq:thm.W2}
 \pr{ \dist \left(\Sigma_\Bx,\Sigma_\By\right) \le s }
\le C\, L^{(2N+1)d} s^{2/3}.
\ee
\end{theorem}

\proof

The proof of Theorem \ref{thm:W2.Vone} will be obtained by collecting the assertions
of Theorem \ref{thm:RCM} and Lemmas \ref{lem:dist.WS} and \ref{lem:main.lemma}.
\vskip2mm

Given a random field $V(x;\om )$, $x\in\cZ$, and a finite subset $\cQ\subset\cZ$, consider
the sample mean and the fluctuations of $V$ relative to $\cQ$:
$$\xi_\cQ(\om) := (\sharp\,\cQ)^{-1}\sum_{x\in\cQ} V(x;\om),\;\;
\eta_x(\om) = \eta_{x,\cQ}(\om) = V(x;\om) - \xi_\cQ(\om),$$
and the sigma-algebra
$\fF_\cQ$ generated by the fluctuations $\{\eta_x, x\in\cQ\}$ and by $\{V(y;\om), y\not\in\cQ\}$.

We use the following property reflecting regularity of the conditional mean:
\vskip3mm

\RCM:
\emph{ There exist constants $C', C'', A', A'', b', b''\in(0,+\infty)$ such that
for any finite subset $\cQ\subset  \cZ$, the conditional distribution  function
$F_\xi( \cdot \,| \fF_{\cQ})$ of the sample mean $\xi_\cQ$
satisfies for all $s\in(0,1)$
\begin{equation}\label{eq:CMxi}
 \;\;
\pr{ \sup_{t\in \DR} \; |F_\xi(t+s\,| \fF_{\cQ}) - F_\xi(t\,| \fF_{\cQ})| \ge C^\prime (\sharp\,\cQ)^{A^\prime} s^{b^\prime} }
\le C^{\prime\prime} (\sharp\,\cQ)^{A^{\prime\prime}} s^{b^{\prime\prime}}.
\end{equation}
} 

Condition \RCM is fulfilled
for an IID Gaussian field, e.g., with zero mean and a unit variance; in this case
the sample mean
is independent of the fluctuations $\eta_\bullet$ and has a normal distribution with
variance $\sigma^2 = (\sharp\,\cQ)^{-1}$. An elementary argument (cf. \cite{C13a})
shows that \RCM also holds for an IID random field with a
uniform marginal distribution.
Moreover, using standard approximation techniques, one can prove the following result:
%
\btm[Cf. Theorem 6 in \cite{C13a}]\label{thm:RCM}
If a random field $x\in\cZ\mapsto V(x;\om )\in\DR$ obeys {\rm\Vone},
then it satisfies property {\rm\RCM} with
$$
C' = 1, \, A' = 1, \, b' = 2/3,
\;\; C'' = (4 R \ovp)^2,\, A'' = 0, \,b'' = 2/3.
$$
\etm

Before we move further, let us introduce some notation. In Eqn \eqref{eq:def.Pi} we
define the support $\Pi\Bx$ of the configuration $\Bx=(x_1,\ldots ,x_N)\in\cZ^N$, the
support $\Pi\bcB(\Bx,L)$ of the
ball $\bcB= \bcB^{(N)}(\Bx,L)$,  and -- given a subset $\cJ\subset\{1,N\}$
--  the partial supports $\Pi_\cJ\Bx$ and $\Pi_\cJ\bcB$:
\be\label{eq:def.Pi}
\begin{array}{c}
\Pi\Bx =\operatornamewithlimits{\cup}\limits_{1\leq i\leq N}\{x_i\}\subset\cZ ,\;\;
\Pi\bcB =\operatornamewithlimits{\cup}\limits_{1\leq j\leq N}\cB (x_j,L) \subset\cZ,\\
\Pi_\cJ\Bx=\operatornamewithlimits{\cup}\limits_{j\in\cJ}\{x_j\}\subset\cZ,\;\;
\Pi_\cJ\bcB =\operatornamewithlimits{\cup}\limits_{j\in\cJ} \cB (x_j,L)\subset\cZ,
\end{array}\ee
with $\Pi_\varnothing \Bx = \varnothing$ (for $\cJ=\varnothing$).

\bde
A  ball $\bcB^{(N)}(\Bx,L)$ is called weakly separated from $\bcB^{(N)}(\By,L)$ if
there exists a single-particle ball $\cB\subset \cZ$,
of diameter $\diam\,\cB \le 2NL$,  and  subsets  $\cJ_1, \cJ_2\subset\{1,\ldots ,N\}$  such that
$\sharp \cJ_1 >\sharp \cJ_2$ (possibly,  with $\cJ_2=\varnothing$) and
\be\label{eq:cond.WS}
\bal
 \Pi_{\cJ_1} \bcB^{(N)} (\Bx,L) \cup \Pi_{\cJ_2} \bcB^{(N)} (\By,L) \;  &\subset \cB,
\\
 \Pi_{\cJ^c_1} \bcB^{(N)} (\By,L) \cup \Pi_{\cJ^c_2} \bcB^{(N)} (\By,L)  &\subset \cZ\setminus \cB.
\eal
\ee
A  pair of balls $\bcB^{(N)} (\Bx,L)$, $\bcB^{(N)} (\By,L)$ is called weakly separated if at least one of the
balls is weakly separated from the other.

To stress the role of the ball $\cB$, we will say, where appropriate, that
$\bcB^{(N)}(\Bx,L)$ and  $\bcB^{(N)}(\By,L)$ are weakly $\cB$-separated.
\ede

\ble\label{lem:dist.WS} {\rm{(Cf. Lemma  2.3 in \cite{C10a})}}
Any pair of $3NL$-distant balls $\bcB^{(N)}(\Bx,L)$,  $\bcB^{(N)}(\By,L)$
is weakly separated.
\ele

The proof of Lemma \ref{lem:dist.WS} repeats that of  Lemma 2.3 in \cite{C10a}
and is omitted.

\begin{lemma}\label{lem:main.lemma}
Let $(x,\om) \to V(x;\om)$ be a random field  satisfying the condition
{\rm\RCM}. Assume that the balls $\bcB^{(N)}(\Bx,L)$, $\bcB^{(N)}(\By,L)$ are weakly separated.
Then for any $s>0$ the following bound holds for the  spectra
$\Sigma_\Bx:=\Sigma(\BH^{(N)}_{\bcB (\Bx,L)})$ and $\Sigma_\By:=\Sigma(\BH^{(N)}_{\bcB (\By,L)})$:
\be\label{eq:LmainL}\begin{array}{l}
\pr{ \dist(\Sigma_\Bx,\Sigma_\By)) \le s }\\
%
\qquad\le \big(\sharp\,\bcB^{(N)} (\Bx,L)\big) \, \big(\sharp\,\bcB^{(N)}(\By,L )\big)\, C'L^{A'} (2s)^{b'}
+ C''L^{A''} (2s)^{b''}\end{array}\ee
where $A^\prime ,A^{\prime\prime},C^\prime ,C^{\prime\prime}b^\prime ,b^{\prime\prime}
\in (0,\infty )$ are as in \eqref{eq:CMxi}.
\end{lemma}

\proof
Let $\cB$ be a ball satisfying the conditions \eqref{eq:cond.WS}
for some $\cJ_1, \cJ_2 \subset \{1,\ldots ,N\}$ with  $\sharp\,\cJ_1=n_1 > n_2 =\sharp\,\cJ_2$.
Introduce the sample mean $\xi=\xi_{\cB}$ of $V$ over $\cB$ and the respective fluctuations
$\{\eta_x := V(x;\om) - \xi_\cB(\om), \, x\in \cB \}$.

Operators $\BH^{(N)}_{\bcB (\Bx,L)}(\omega)$,  $\BH^{(N)}_{\bcB (\By,L)}(\omega)$ read as follows:
\begin{equation}\label{eq:Ham.decomp}
\BH^{(N)}_{\bcB (\Bx,L)}(\omega) = n_1 \xi(\omega) \, \B{I} + \BA'(\omega), \;
\BH^{(N)}_{\bcB (\By,L)}(\omega) = n_2 \xi(\omega) \,\B{I} + \BA''(\omega)
\end{equation}
where operators $\BA'(\omega)$ and $\BA''(\omega)$ are $\fF_{\cB}$-measurable. Let
$$\Sigma_\Bx=\{ \lambda_1, \ldots, \lambda_{K^\prime}\}\;\hbox{ and }\;
\Sigma_\By=\{ \mu_1, \ldots, \mu_{K^{\prime\prime}}\},$$
where $K^\prime =\sharp\,\bcB (\Bx, L)$ and
$K^{\prime\prime}= \sharp \bcB (\By,L)$.

Owing to \eqref{eq:Ham.decomp}, we have
$\lambda_j(\omega) = n_1\xi(\omega) + \lambda_j^{(0)}(\omega)$,
$\mu_j(\omega) = n_2\xi(\omega) + \mu_j^{(0)}(\omega)$,
where the random variables
$\lambda_j^{(0)}(\omega)$ and $\mu_j^{(0)}(\omega)$ are $\fF_{\cB}$-measurable. Therefore,
$$
\lambda_i(\omega) - \mu_j(\omega) =  (n_1-n_2)\xi(\omega) + (\lambda_j^{(0)}(\omega) -  \mu_j^{(0)}(\omega)),
$$
with $n_1-n_2 \ge 1$, owing to our assumption.
Further, we can write
$$
\bal
\pr{ \dist(\Sigma_\Bx, \Sigma_\By)) \le s }
\le \sum_{1 \le i \le K^\prime} \; \sum_{1 \le j \le K^{\prime\prime}} \Bigesm{ \pr{ |\lambda_i - \mu_j| \le s \,| \fF_{\cB}}}.
\eal
$$
Note that for all $i$ and $j$ we have
$$
\bal
\pr{ |\lambda_i - \mu_j| \le s \,|\, \fF_{\cB}}
& = \pr{ |(n_1 - n_2)\xi + \lambda_i^{(0)} - \mu_j^{(0)}| \le s \,| \fF_{\cB}}
\\
&\le \displaystyle \mnu_L( 2|n_1 - n_2|^{-1} s \,|\, \fF_{\cB}) \le \mnu_L( 2 s \,| \fF_{\cB}).
\eal
$$
Set
$$
\cD_L = \bigl\{\om: \sup_{t\in\DR} \;
\big|F_\xi(t+s\,| \fF_{\cB}) - F_\xi(t\,| \fF_{\cB})\big|\ge C'L^{A'} s^{b'} \bigr\}.
$$
By \RCM, $\pr{\cD_L} \le C''L^{A''} s^{b''}\}$.
Therefore, denoting $\cD^\complement_L = \Om\setminus \cD_L$,
\be\label{eq:proof.W2.RCM.4}
\bal
\pr{ \dist(\Sigma_\Bx, \Sigma_\By) \le s }
&\le \esm{ \one_{\cD^\complement_L} \pr{ \dist(\Sigma_\Bx, \Sigma_\By) \le s \,| \fF_{\cB}}}
+ \pr{\cD_L}
\\
&\le (\sharp\,\bcB(\Bx,L)) \cdot (\sharp\,\bcB (\By,L))\, C'L^{A'} s^{b'}
+ C''L^{A''} s^{b''} ,
\eal
\ee
as claimed in \eqref{eq:LmainL}. This finishes the proof of Lemma \ref{lem:main.lemma}.
\qedhere

\vskip3mm

By Theorem \ref{thm:RCM}, property \RCM is fulfilled with
$b'=b''=2/3$, $A'=1$, $A''=0$. Hence, the RHS in \eqref{eq:proof.W2.RCM.4}
is bounded by ${\ov C}L^{Nd} s^{2/3}$.
This completes the proof of Theorem \ref{thm:W2.Vone}.\qedhere

\vskip3mm

Theorem \ref{thm:W2.Vone}
is essential in the proof of Theorem \ref{thm:2vol.VEMSA.ETV}. Namely,
it allows us to infer from the fixed-energy
decay bounds (which are simpler to establish) their energy-interval counterparts, required for the proof
of spectral and dynamical localization, without an additional scaling analysis employed in the bootstrap
multi-scale approach (cf. \cite{KN13a}, \cite{GK01}).

\subsection{Weakly interactive balls}

\begin{definition}\label{def:WI}
An $N$-particle ball $\bcB^{(N)} (\Bu,L)$, with $N\ge 2$,
centered at $\Bu =(u_1,\ldots ,u_n)$, $u_i\in\cZ$,
is called weakly interactive ({\rm{WI}}) if
\be
\diam( \Pi \Bu):=\max_{1\leq i<j\leq N} \rd (u_i,u_j) > 3 NL,
\ee
and strongly interactive ({\rm{SI}}), otherwise.
\end{definition}

The meaning of Definition \ref{def:WI} is that a particle system in a WI ball can be
decomposed into distant subsystems that interact ``weakly`` with each other, whereas
for an SI ball such a decomposition is not possible. See Lemma \ref{lem:WI.decomp}.

\ble\label{lem:WI.decomp}
For any {\rm{WI}} ball $\bcB^{(N)}(\Bu, L)$ there exists a decomposition
$\{1, \ldots, N\}$ $= \cJ \cup \cJ^\rc$, with $\cJ^\rc := \{1, \ldots, N\}\setminus\cJ$,
such that,
\be\label{eq:WIdecomp}
\rd\left( \Pi_\cJ \bcB^{(N)}(\Bu,L),  \Pi_{\cJ^\rc} \bcB^{(N)}(\Bu,L)\right) > L.
\ee
\ele

\begin{proof}
Suppose that $\diam( \Pi \Bu) > 3 N L$; we want to
show that the projection $\Pi \bcB(\Bu,3L/2)$ is a disconnected subset of $\cZ$.

Assume otherwise; then every partial projection $\Pi_\cJ\Bu$,
$\varnothing\subset\cJ\subset\{1,\ldots ,N\}$,
is at distance $\le 2 \cdot \frac{3L}{2}=3L$ from $\Pi_{\cJ^\rc}\Bu$. Then a straightforward
induction in $N\ge 2$ shows that
$\diam\; \Pi\Bu \le (N-1) \cdot 3L < 3NL$, contrary to our hypothesis.

Now, as $\Pi\bcB (\Bu,3L/2)$ is disconnected, there exists a nontrivial decomposition
$\{1, \ldots, N\} = \cJ \cup \cJ^\rc$ for which
$$\begin{array}{l}
\rd\left( \Pi_\cJ \bcB (\Bu,3L/2),  \Pi_{\cJ^\rc} \bcB (\Bu,3L/2)\right)\ge 1\\
\qquad{}\Rightarrow
\rd \left( \Pi_\cJ \bcB (\Bu,L),  \Pi_{\cJ^\rc}\bcB (\Bu,L)\right)
> {\textstyle \frac{1}{2}} L + {\textstyle\frac{1}{2}} L = L,\end{array}
$$
as asserted in Eqn \eqref{eq:WIdecomp}.
\end{proof}

The decomposition $(\cJ,\cJ^{\rc})$ figuring in Lemma \ref{lem:WI.decomp} may be
not unique. We will assume that such a decomposition (referred to as
the canonical one) is associated in some unique way with every $N$-particle WI ball.
Accordingly, we fix the notation $N^\prime =\sharp\cJ$, $N^{\prime\prime}=\sharp\cJ^{\rc}
=N-N^\prime$, and further -- for $\Bx =(x_1,\ldots ,x_N)\in\bcB (\Bu,L)$ --
\be\label{eq:candcomp}\begin{array}{l}
\Bx_\cJ=(x_{i_1},\ldots ,x_{i_{N^\prime}}),\; \Bx_{\cJ^\rc}=(x_{j_1},\ldots ,x_{j_{N^{\prime\prime}}})\\
\hbox{where}\;\;
\cJ=\{i_1,\ldots ,i_{N^\prime}\},\;\;\cJ^\rc=\{j_1,\ldots ,j_{N^{\prime\prime}}\},\\
\hbox{with $1\leq i_1<\ldots <i_{N^\prime}\leq N$,
$1\leq j_1<\ldots <j_{N^{\prime\prime}}\leq N$.}\end{array}\ee
This  gives rise to the Cartesian product representation
\be\label{eq:can.decomp}
\begin{array}{c} \bcB =\bcB^\prime\times\bcB^{\prime\prime}\;\hbox{ where } \bcB=\bcB^{(N)}(\Bu,L)
\hbox{ and}\\
\bcB^\prime = \bcB^{(N^\prime)}(\Bu_\cJ,L),\;\bcB^{\prime\prime}=
\bcB^{(N^{\prime\prime})}(\Bu_{\cJ^\rc},L),\end{array}\ee
which we also call the canonical factorization.

Consequently, the operator $\BH^{(N)}_{\bcB}$ in
a WI ball $\bcB=\bcB (\Bu,L)$ can be represented in the following way:
\be\label{eq:Ham.decomposable}
\BH^{(N)}_{\bcB} = \BH^{(N^\prime)}_{\bcB^\prime}\otimes \BI^{(N^{\prime\prime})}
+ \BI^{(N^\prime)} \otimes \, \BH^{(N^{\prime\prime})}_{\bcB^{\prime\prime}}+
\BU_{\bcB^\prime ,\bcB^{\prime\prime}}.\ee
Here the summand $\BU_{\bcB^\prime ,\bcB^{\prime\prime}}$
takes into account the interaction between subsystems in balls $\bcB^\prime$ and
$\bcB^{\prime\prime}$ and has a small norm for $L$ large. Operators $\BH^{(N^\prime)}_{\bcB^\prime}$
and $\BH^{(N^{\prime\prime})}_{\bcB^{\prime\prime}}$ are called the reduced Hamiltonians
(for the WI ball $\bcB$).

\ble\label{lem:SI.full.sep}
Let $\bcB^{(N)}(\Bx, L)$, $\bcB^{(N)}(\By, L)$ be a pair of \SI balls with
$\brho(\Bx,\By) > 8 NL$. Then
\be\label{eq:dsit.SI.indep}
\Pi \bcB^{(N)}(\Bx, L) \cap \Pi \bcB^{(N)}(\By, L) = \varnothing,
\ee
and, consequently, the random operators $\BH_{\bcB^{(N)}(\Bx, L)}(\om)$ and
$\BH_{\bcB^{(N)}(\By, L)}(\om)$ are independent.
\ele

\begin{proof}
By definition, for any \SI balls $\bcB^{(N)}(\Bx, L)$, $\bcB^{(N)}(\By, L)$
we have
$$
\max_{i,j} \rd(x_i, x_j) \le 3NL, \;\; \max_{i,j} \rd(y_i, y_j) \le 3NL,
$$
and it follows from the assumption $\brho(\Bx,\By) > \rA NL$ that
for some $i',j'\in\{1, \ldots, N\}$
$
\rd(x_{i'}, y_{j'}) > 3NL,
$
thus for any $i,j\in\{1, \ldots, N\}$
$$
\bal
\rd(x_i, y_j) \ge \rd(x_{i'}, y_{j'}) - \rd(x_{i'}, x_{i'}) - \rd(x_{j'}, y_{j})
> 8NL - 6NL =2NL.
\eal
$$
Therefore, with $N\ge 1$,
$$
\dist\big(\Pi \bcB_L(\Bx), \Pi \bcB_L(\By) \big) > 2(N-1) L \ge 0,
$$
so $\Pi \bcB^{(N)}(\Bx, L) \cap \Pi \bcB^{(N)}(\By, L) = \varnothing$.
Consequently, the samples of the random potential in   $\BH_{\bcB^{(N)}(\Bx, L)}(\om)$
and $\BH_{\bcB^{(N)}(\By, L)}(\om)$ are independent.
\end{proof}

Throughout the paper we consider  a sequence of integers $L_k>1$ of one of the two forms
\be\label{eq:def.Lk}
{\rm{(a)}}\quad L_{k+1} := L_k \rB  ,\; k=0,1,\ldots ,\;\hbox{ or (b)}\quad
L_{k+1} = \left\lfloor L_k^\alpha\right\rfloor ,\; k=0,1,\ldots \ee
with given initial positive integer values $L_0, \rB$ and a scaling exponent $\alpha >1$.
Referring to \eqref{eq:def.Lk}, we consider

\bde\label{def:CNR}
A ball $\bcB^{(N)} (\Bx ,L_k)$, $k\ge 1$,
is called $(E,\beta )$-completely non-resonant ({\rm $(E,\beta )$-CNR}) if
all concentric balls $\bcB^{(N)}(\Bx ,\ell)$ with $L_{k-1}\le\ell\le L_k$
are {\rm$(E,\beta )$-NR}.\ede

The next definition is based upon Definitions  \ref{def:NR.NS} and \ref{def:CNR}. Here we use
parameter $m_N$ of the following form:
\be\label{eq:def.mN}
m_N := m^*\,\big(1 + 3L_0^{-\delta+\beta}\big)^{N^*-N+1} \ee
with some $\beta\in (0,1)$, $\delta\in (\beta ,1)$ and $m^*>0$.
Note that for all $N=1, \ldots, N^*-1$, we have (cf. \eqref{eq:table1})
$$
m_N = \big(1 + 4L_0^{-\delta+\beta}\big) m_{N+1} > m_{N+1} \ge m_*.
$$

\bde\label{def:FNR}
Assume that $\beta\in (0,1)$, $\delta\in (0,1]$ and
$\nu_N$ are as in \eqref{eq:table1}. Next, let $\bcB =\bcB^{(N)}(\Bu ,L)$ be a  {\rm WI} ball with the
canonical factorization \eqref{eq:can.decomp} and suppose that, for a given
$E\in\DR$, $\bcB$ is {\rm\EbNR}. The following properties
{\rm{(i)}}, {\rm{(ii)}} are defined in terns of the reduced Hamiltonians
$\BH^{(N^\prime)}_{\bcB^\prime}$ and $\BH^{(N^{\prime\prime})}_{\bcB^{\prime\prime}}$.
We say that

{\rm{(i)}} $\bcB$  is $(E,\beta )$-fully non-resonant ($(E,\beta )$-{\rm{FNR}}) if
\be\label{eq:CNRJ}\begin{array}{l}
\hbox{$\forall$ $\lam^\prime \in\Sigma\left(\BH^{(N^\prime)}_{\bcB^\prime}\right)$,  ball
$\bcB^{\prime\prime}$ is $(E-\lam^\prime,\beta)${\rm-CNR}}\\
\hbox{and}\\
\hbox{$\forall$ $\lam^{\prime\prime} \in\Sigma\left(\BH^{(N^{\prime\prime})}_{\bcB^{\prime\prime}}\right)$,
 ball  $\bcB^\prime$ is $(E-\lam^{\prime\prime},\beta )${\rm-CNR}.}\end{array}\ee

Furthermore, we say that

{\rm{(ii)}} $\bcB$  is $(E,\delta, \nu_{N^\prime},\nu_{N^{\prime\prime}})$-partially non-singular
($(E,\delta, m_{N^\prime},m_{N^{\prime\prime}})$-{\rm{PNS}})
if in Eqns \eqref{eq:CNRJ} we have properties
$(E-\lam^\prime,\delta, m_{N^\prime})$-{\rm{NS}} and $(E-\lam^{\prime\prime},\delta,
m_{N^{\prime\prime}})$-{\rm{NS}}, instead of $(E-\lam^\prime,\beta)${\rm{-CNR}} and
$(E-\lam^{\prime\prime},\beta )${\rm{-CNR}}, respectively:

\be\label{eq:PNS}
\begin{array}{l}
\hbox{$\forall$ $\lam^\prime \in\Sigma\left(\BH^{(N^\prime)}_{\bcB^\prime}\right)$,  ball
$\bcB^{\prime\prime}$ is $(E-\lam^\prime,\delta, m_{N^\prime})$-{\rm{NS}}}\\
\hbox{and}\\
\hbox{$\forall$ $\lam^{\prime\prime} \in\Sigma\left(\BH^{(N^{\prime\prime})}_{\bcB^{\prime\prime}}\right)$,
 ball  $\bcB^\prime$ is $(E-\lam^{\prime\prime},\delta,
m_{N^{\prime\prime}})$-{\rm{NS}}.}
\end{array}
\ee
\ede

Property $(E,\delta, m_{N^\prime},m_{N^{\prime\prime}})$-{\rm{PNS}} is employed in
Lemma \ref{lem:WITRONS.subexp} whereas $(E,\beta )$-{\rm{FNR}} in Theorem \ref{thm:FNR}
(and in several places later).
Furthermore, in Theorem \ref{thm:FNR} we refer to the case (a) in Eqn \eqref{eq:def.Lk}:
\be\label{eq:def.Lkexp}
L_{k+1} :=L_0 \rB^k,\;\; k=0,1,\ldots ,\;\hbox{ where }\;L_0,B\in\DN^*, \, B\ge 2.
\ee

\btm\label{thm:FNR}
Fix $b,\beta\in (0,1)$ and $\rB \ge 2$. If $L_0$ is chosen large enough then
for all $E\in\DR$ and $k\ge 0$,
\be
\bal
\pr{ \hbox{{\rm{ball}} $\bcB^{(N)}(\Bu,L_k)$ {\rm{is WI but not $(E,\beta )$-FNR} }}}
\le  2\eu^{ -(2\rB)^{-1} L_k^{\beta b} }.
\eal
\ee
\etm

\proof Let ball $\bcB^{(N)}(\Bu,L_k)$ be WI and  \EbNR. To shorten the notation, set
-- referring to canonical factorization \eqref{eq:can.decomp} --
\be\label{BprimeBprimeprime}
\Sigma'=\Sigma\left(\BH^{(N^\prime)}_{\bcB^\prime}\right)\;\hbox{ and }\;
\Sigma''=\Sigma\left(\BH^{(N^{\prime\prime})}_{\bcB^{\prime\prime}} \right).\ee
Consider the event
$\cS = \{\bcB \text{ is not $(E,\beta )$-FNR}\}$. Then
$\cS \subset \cS' \cup \cS''$, where
\be\label{SprimeSprimeprime}
\bal
\cS' &= \left\{ \exists\, \lam''\in\Sigma^{\prime\prime}:
  \bcB^\prime\text{ is not $(E-\lam^{\prime\prime},\beta)$-CNR} \right\},
\\
\cS'' &= \left\{ \exists\, \lam'\in\Sigma^\prime :
  \bcB^{\prime\prime} \text{ is not $(E-\lam^\prime ,\beta)$-CNR}  \right\}.
\eal\ee
First, assess $\pr{\cS^\prime}$. Denoting by $\fF^{\prime\prime}$ the sigma-algebra generated
by the values $V(x)$, $x\in\Pi_{\cJ^\rc}\bcB$, write:
\be\label{eq:estprob}\bal
\pr{ \cS' } &\le(\sharp\;\bcB^{\prime\prime}\,)\,  \sum_{L_{k-1} \le \ell \le L_{k} }
\max_{\lam''\in\Sigma''}
\Bigesm{ \pr{ \bcB^\prime  \text{ is $(E-\lam'' ,\beta)$-R} \,|\, \fF''}   }
\\
& \le L_{k} \cdot\left(C_\cZ^N L_k^{Nd}\right) \; \sup_{E^\prime\in\DR}\;
    \pr{ \bcB^{(N^\prime )}(\Bu_\cJ,\ell )  \text{ is $(E^\prime ,\beta)$-R}}\\
&\le C_\cZ^N L_k^{Nd+1} \, \eu^{ - L_{k-1}^{\beta b}}
 = \eu^{  - \rB^{-\beta b} L_{k}^{\beta b} + \ln\left(C_\cZ^N L_k^{Nd+1} \right)}
\le \eu^{ -(2\rB )^{-1} L_k^{\beta b} },
\eal\ee
provided that $L_0$ is large enough, which we assumed.

As the roles of $\bcB^\prime$ and $\bcB^{\prime\prime}$are  symmetric, the same upper
bound holds  for $\pr{\cS^{\prime\prime}}$. This completes the proof of Theorem \ref{thm:FNR}.
\qedhere

\vskip1mm

Theorem \ref{thm:FNR} will be instrumental for the proof of Theorem
\ref{cor:prob.WI.S.large.g.DR} and -- in a form modified to case (b) in \eqref{eq:def.Lk}
-- in the proof of Theorem \ref{cor:prob.WI.S.exp}.

\section{Fixed- and variable-energy estimates }  
\label{Sec:FEMSA.subexp}

The aim in this section is to prove assertion {\textbf{(A)}} of Theorem
\ref{thm:Main.DL.Vone.Uone.strong.g}; the main
technical tool is provided by so-called variable-energy estimates.
(In \cite{KN13a} the term continuum-energy has been used.) An
example is Theorem \ref{thm:2vol.VEMSA.ETV}. In the MPMSA, such
bounds are difficult to obtain;
in a sense, they represent a bottleneck of the whole method.
Nevertheless, until subsection \ref{ssec:FEMSA.to.VEMSA}
 we work with much simpler fixed-energy estimates, preparing the grounds
 for the passage to the variable-energy ones. The sequence $\{L_k\}$ is taken
 in this section of the form \eqref{eq:def.Lkexp}.

Throughout the section we use, in various combinations,
inequalities listed in \eqref{eq:table1}. These inequalities are imposed upon
key parameters of the inductive schemes involved. Namely,
we employ the following parameters: (i) $\rB$ and $L_0$ (positive integers); (ii)
$\kappa\in (0,\zeta ]$ (bounds the decay of the EFCs); (iii) $\beta\in (0,1)$ (a resonance/nonresonance
threshold value, emerging in \eqref{eq:def.NR}); (iv) $m^*\geq 1$ giving rise to a `mass` $m_N$ and $\delta\in (0,1)$ (a sub-exponential decay parameter figuring in \eqref{eq:def.NS}); (v) $K$ (a nonnegative integer appearing in
\eqref{eq:def.K} and controlling
the number of singular balls of radius $L_k$ inside a ball of radius $L_{k+1}$); (vi) $\nu^*\geq 1$ used in
\eqref{eq:SSIsexp} through
the scaled value $\nu_N$ controlling the decay of the so-called  singularity
probability. In the table \eqref{eq:table1} we show the relations between these parameters.
(A specific form of some of these relations is chosen for technical convenience.) Recall, $N$ takes
values $1,...,N^*$.

The integer $K$ appears in Definition \ref{def:K.good.subexp} below
and also in Definiiton \ref{def:K.good.exp} (cf. Sect.~\ref{Sec:exp.decay.EF}).
In Sect.~\ref{Sec:exp.decay.EF}, the MSA induction is adapted to the length scale sequence
satisfying $L_{k+1}=\lfloor L_k^\alpha \rfloor$, where $\alpha>1$ depends upon the decay exponent $\zeta>0$
of the interaction potential, and $K$ is to be chosen large enough, depending upon $\zeta$.
In Sect.~\ref{Sec:FEMSA.subexp}, it suffices to set $K=1$ to obtain sub-exponential decay of EF correlators
with some exponent $\kappa>0$, but we keep the value of $K$ in symbolic form. It is worth mentioning that by
choosing $K>1$ large enough, one can make $\kappa\in(0,1)$ arbitrarily close to $1$, but this requires some additional
analysis which we omit for brevity and clarity of presentation. For further details, see the work by Klein and Nguyen
\cite{KN13a}, adapting to the multi-particle setting the bootstrap MSA techniques, originally developed in \cite{GK01}.

\renewcommand{\arraystretch}{1.7}
\be\label{eq:table1}\hbox{\begin{tabular}{|l|l|}
  \hline
  $0 < \kappa <\zeta,\;0 < \beta < \delta < \zeta\wedge 1\begin{matrix}\;
  \\ \;\end{matrix}$ &  $\beta +\diy\frac{\ln(8\rB) }{\ln L_0}<\delta < 1 - \frac{\ln 12}{\ln\rB} $
 \\
  \hline
$m^*,\nu^*\geq 1$ &
$\begin{array}{l}\hbox{$\rB \ge 24 N^* K$;
  $L_0$ large enough}\\
\hbox{depending on $\beta ,\delta ,\rB,K,m^*,\nu^*$} \end{array}$ \\
  \hline
  $m_N = m^*\,\big(1 + 4L_0^{-\delta+\beta}\big)^{N^*-N+1}$ & $\nu_N = \nu^*\, (2 \rB^\kappa)^{N^*-N+1}$
\\
\hline
\end{tabular}}\ee

For definiteness, we assume \eqref{eq:table1} to be satisfied throughout the whole section
\ref{Sec:FEMSA.subexp}, regardless of whether a particular parameter is involved in a given
assertion or not. This will not be reminded every time again, although basic ranges for values
of $\beta,\delta,\kappa, m^*,\nu^*$ will be outlined. (A number of technical statements remain
valid under broader restrictions than those from \eqref{eq:table1}.)

\ssect{Scaling the GFs. Property \SSI{N,k}}\label{propertySSI}

\bde
\label{def:K.good.subexp}
Suppose that the following values are given: $E\in\DR$, $\beta ,\delta\in(0,1)$,
$m^*\geq 1$ and integers $k, K \ge 0$.
An $N$-particle ball $\bcB=\bcB^{(N)}(\Bu,L_{k+1})$ is called $(E,m_N,K)$-good
($(E,m_N,K)$-{\rm G}) if  $\bcB$ is $(E,\beta )$-{\rm{CNR}} (cf. Definition {\rm\ref{def:CNR}}) and
\be\label{eq:def.K}\begin{array}{r}
\hbox{$\bcB$ contains no collection of $\;\ge K+1$ balls of radius $L_k$}\quad{}\\
\hbox{which are pairwise $8NL_k$-distant and \mEdmS{\delta,m_N}.}\end{array}\ee
\ede

In this definition we omitted parameters $\delta,\beta\in (0,1)$ from the notation\\
$(E,m_N,K)$-{\rm G}).

\begin{lemma}\label{lem:NR.NT.implies.NS}
Given $\beta ,\delta,m^*,K$ and $\rB$,  suppose that
$L_0$ is large enough.
If a ball $\bcB=\bcB^{(N)}(\Bu,L_{k+1})$ is {\rm\EbNR} and $(E,m_N,K)$-{\rm G},
then $\bcB$  is $(E,\delta ,m_N)$-{\rm{NS}}.
The assertion remains valid under a weaker condition than {\rm\EbNR}:
\be\label{eq:lem.NR.NT.implies.NS}
\dist\left( \Sigma\left(\BH^{(N)}_\bcB)\right), E\right) \ge \eu^{-L^\beta}.
\ee
\end{lemma}

\proof
In this proof we use Lemmas  \ref{lem:rad.many.singular} and
\ref{thm:GF.is.ell.q.Xi.dominated.subexp}  from Appendix \ref{Sec:Ddecay}.
Fix $K^\prime\le K$ and a maximal collection of $K'$ pairwise $8NL_k$-distant, $(E,\delta ,m_N)$-NS
balls of radius $L_k$ lying in $\bcB$. Let $\bcN$ denote the $L_k$-neighborhood
of the union of these balls. Then any ball $\bcB^{(N)}(\Bv,L_k)$ with $\Bv\in\bcB\setminus \bcN$
is $(E,\beta ,m_N)$-NS. Bearing in mind Lemma \ref{thm:GF.is.ell.q.Xi.dominated.subexp},
denote by $\Xi$ be the union of all spherical layers $\bcL_r(\Bu)$ such that
$\bcL_r(\Bu)\cap\bcN\ne \varnothing$.
It follows from table \eqref{eq:table1} (the relations between $\delta$ and $\beta$) that
\be
m_N - 2 L_k^{-\delta} L_{k+1}^\beta = m_N \left(1 - 2 m_N^{-1} L_k^{-\delta+\beta} \rB^\beta \right)
\ge {\textstyle\frac{3}{4}} m_N >0.
\ee
Thus, by Lemma \ref{thm:GF.is.ell.q.Xi.dominated.subexp}, the function
$$
f: \Bx\in\bcB\mapsto\left|G^{(N)}_{\bcB}(\Bu, \Bx;E)\right|
$$
is $(\ell,q,\Xi)$-dominated in $\bcB$,
with $q \le \eu^{- \frac{3}{4} m_N L_k^\delta}$. Cf. Eqn \eqref{functionf}.

Owing to Lemma \ref{lem:rad.many.singular}, we can write,
with the convention $-\ln 0 = +\infty$, that $-\ln f(\Bx)$ is bounded below by
$$
- \ln\left\{
\eu^{L_{k+1}^\beta}\exp\,\left[-\frac{3m_N}{4}L_k^\delta\cdot
\frac{L_{k+1} - (8NK+2) L_k - 2L_k}{ L_k + 1}\right]\right\}
$$
thus by virtue of conditions in Eqn \eqref{eq:table1} (in particular, with
$L_0\ge 3$, $B \ge 24N^*K\ge 24NK$ and $\frac{1}{4}B^{1 - \delta}\ge 3$),
one obtains by a simple calculation
$$
-\ln f(\Bx) \ge 2 m_NL_{k+1}^\delta \ge m_NL_{k+1}^\delta
+ \ln(C_\cZ^{2N} L_{k+1}^d).$$
\qedhere

Lemma \ref{lem:NR.NT.implies.NS} has a multiple use: it is  needed in the proof of Theorems
\ref{cor:prob.WI.S.large.g.DR} and \ref{thm:MSA.main.fixed}.

Given $L_0,\rB,\delta , \kappa ,m^*,\nu^*$, consider the following  property \SSI{N,k} depending upon
$N$ and $k$ ({\textsf S} stands for singularity):

\par
\vskip1mm\noindent
\SSI{N,k}: \qquad $\forall\, E\in\D{R}$,\; $1\leq n\leq N$ and configuration $\Bu\in\bcZ$
\be\label{eq:SSIsexp}
\pr{ \text{ball $\bcB^{(n)}(\Bu,L_k)$ is $(E,\delta,m_n)$-S }  }
\le
\eu^{ - \mnu_n L_{k}^\kappa } .
\ee
The MPMSA inductive scheme consists in checking \SSI{N,k} $\forall$ $N$ and $k$.
The initial step of induction in $k$ is established in Theorem \ref{thm:L0.large.g}.



\btm\label{thm:L0.large.g} Suppose a positive integer $M$
and an $M\times M$ Hermitian matrix $\BA$  are given, as well as
random variables (not assumed to be independent)
$W_1$, $\ldots$, $W_M$, with continuous distribution
functions $F_{W_i}$, $1\le i \le M$.
Let $\BW(\om)$ be the diagonal random matrix ${\rm{diag}}(W_1(\om), \ldots, W_M(\om))$.
For any $s>0$ and $\eps\in(0,1)$, there exists
$g^*<\infty$ such that if $|g|\ge g^*$ then
$$
\sup_{E\in \DR} \; \pr{ \| \big(\BA + g\BW - E\BI\big)^{-1} \| > s  } \le \eps < 1.
$$

Consequently, $\forall$ $\delta\in (0,1)$, $\kappa\in (0,\zeta )$  and $m^*,\nu^*\geq 1$,
$\exists$ $g_0\in (0,\infty )$ such that $\forall$ $1\leq N\leq N^*$ and positive integer $L_0$,
property {\rm{\SSI{N,0}}} holds true.
\etm

The proof is omitted; it is based on a well-known argument employed in a number
of papers on the MSA (cf., e.g., \cite{DK89}*{Proposition A.1.2}) and is not contingent
upon the single- or multi-particle structure of the random diagonal entries of the matrix $\BA$.

\ssect{The GFs in WI balls. The MPMSA induction}\label{ssec:GreeninWI} 

\begin{lemma}\label{lem:WITRONS.subexp}
Fix $\beta ,\delta\in (0,1)$, $m^*\geq 1$ and an energy value $E\in\DR$.
Consider a {\rm WI} $N$-particle ball $\bcBN (\Bu,L_k)$ with a canonical factorization
$\bcB^{(N^\prime)}(\Bu_\cJ,L_k) \times \bcB^{(N^{\prime\prime})}_L(\Bu_{\cJ^\rc},L_k)$.
Suppose that
$\bcB^{(N)}(\Bu, L_k)$ is {\rm\EbNR} and $(E,\delta, m_{N^\prime},m_{N^{\prime\prime}})$-{\rm{PNS}}.
Cf. Definition {\rm{\ref{def:FNR}, (ii)}}.
If $L_0$ is large enough
then  $\bcB^{(N)}(\Bu,L_k)$ is {\rm$(E, \delta, m_{N})${\rm-NS}}.
\end{lemma}
\proof See Appendix \ref{ssect:AppendixA1}. \qedhere

Lemma \ref{lem:WITRONS.subexp} is used in the proof of Theorem
\ref{cor:prob.WI.S.large.g.DR}.

\bthm\label{cor:prob.WI.S.large.g.DR}
Assume property {\rm\SSI{N-1,k}} for some given $L_0,\rB>1$, $\delta\in (0,1)$,
$\kappa\in (0,\zeta)$  and $m^*,\nu^*\ge 1$
(see Eqn {\rm{\eqref{eq:SSIsexp}}}).
Then if $L_0$ is large enough then for any $E\in \DR$ and {\rm WI} ball $\bcB^{(N)}(\Bu,L_k)$,
\be\label{eq:lem.WI.S}
\pr{\text{ $\bcB^{(N)}(\Bu,L_k)$  is   {\rm\mEdmS{\delta,m_N}} }}
\le 2 \eu^{-\frac{3}{2} \nu_N L_{k+1}^\kappa}.
\ee
Consequently, for $L_0$ large enough, $\forall$ $\Bx\in\bcZ$,
\be\label{eq:prob.S.k+1}
\begin{array}{r}
\pr{ \bcB^{(N)}(\Bx,L_{k+1})
\text{ contains a {\rm WI} {\mEdmS{\delta,m_N}} ball $\bcB^{(N)}(\Bu,L_k)$}}
\quad{}\\
\qquad
\le C_\cZ^N L_{k+1}^{Nd} \cdot
2 \eu^{- \frac{3}{2} \nu_N L_{k+1}^\kappa}
\le  \diy\frac{1}{4 }  \eu^{- \nu_N L_{k+1}^\kappa}.
\end{array}\ee
\ethm

\proof
Denote by $\cS$ the event in the LHS of \eqref{eq:lem.WI.S}. Set $\bcB=\bcB^{(N)}(\Bu,L_k)$
and write the canonical factorization $\bcB=\bcB^\prime\times\bcB^{\prime\prime}$ with
reduced operators $\BH^\prime=\BH^{(N^\prime)}_{\bcB^\prime}$ and $\BH^{\prime\prime}=\BH^{(N^{\prime\prime})}_{\bcB^{\prime\prime}}$
(cf. \eqref{eq:can.decomp}, \eqref{eq:Ham.decomposable}).
By  Lemma \ref{lem:NR.NT.implies.NS},
\be\label{eq:proof.lem.WI.T.a}
\bal
\pr{\cS } & < \pr{\text{ $\bcB$  is not  \FNR }}
\\
&
+ \pr{\text{ $\bcB$  is  \FNR and \mdmS{E,\delta,m_N} }}.
\eal
\ee
The first term in the RHS is assessed in Theorem \ref{thm:FNR},
so we focus on the second summand.
Apply Lemma \ref{lem:WITRONS.subexp} and introduce events $\cS^\prime$ and
$\cS^{\prime\prime}$ by following the framework of Eqn \eqref{SprimeSprimeprime}
and \eqref{eq:estprob}. Then, with $m^{\prime\prime}=m_{N^{\prime\prime}}$,
$$\bal\pr{\cS^\prime }
=\esm{ \pr{
 \exists\,\lam''\in\Sigma(\BH^{\prime\prime}):\, \bcB^\prime
   \text{ is  \mdmS{E-\lam'',\delta,m^{\prime\prime}}} \,\Big|\, \fF'' }}\,.\eal$$

By definition of the canonical decomposition, $
 \PPi \,\bcB^\prime \cap  \PPi \,\bcB^{\prime\prime} =\varnothing$,
and since the random field $V$ is IID, for any $E''\in \DR$, including
$E-\lam''$, the conditional probability  does not depend on the condition:
\be\label{eq:use.mixing}
\pr{\text{ $\bcB^\prime$  is $(E'',\delta ,m^{\prime\prime})${\rm-S}} \,\big|\, \fF''}
\truc{\,=\,}{\rm{a.s.}}{}
\pr{\text{ $\bcB^\prime$  is $(E'',\delta,m^{\prime\prime})${\rm-S}} }.
\ee
On the other hand, by virtue of the (assumed) property \SSI{N-1,k}, for $N^\prime\le N-1$,
\par
\be\label{eq:proj.S.prob}
\pr{ \text{  $\bcB^\prime$  is $(E'',\delta,m^{\prime\prime})${\rm-S}} }
\le \eu^{ - \nu_{N-1} L_k^\kappa } 
= \eu^{ - 2\nu_{N} L_{k+1}^\kappa }.
\ee
Thus, in analogy with \eqref{eq:estprob}, we obtain that
\be\label{eq:proof.lem.WI.T.2a}
\bal \pr{\cS' }
& \le \sharp\;\bcB^{\prime\prime} \, \sup_{E''\in\DR}
\pr{ \text{$\bcB^\prime$  is $(E'',\delta ,m^{\prime\prime})${\rm-S}} }\\
&\qquad \le C_\cZ^N  L_k^{Nd}\, \exp\left\{ - 2 \nu_{N} L_{k+1}^\kappa \right\}
\le \exp\left\{  - {\textstyle \frac{3}{2}} \nu_N L_{k+1}^\kappa \right\};
\eal \ee
here the last inequality  holds for $L_0$ large enough.
Similarly, with $ m^\prime = m_{N^\prime}$,
\be\label{eq:proof.lem.WI.T.3b}\begin{array}{r}
\pr{\cS^{\prime\prime} }
=\esm{ \pr{
 \exists\,\lam^\prime\in\Sigma(\BH^\prime ):\, \bcB^{\prime\prime}
   \text{ is  \mdmS{E-\lam^\prime,\delta,m^\prime}} \,\Big|\, \fF^\prime }}\quad{}\\
\le  \exp\left\{  - {\textstyle \frac{3}{2}} \nu_N L_{k+1}^\kappa \right\}.\end{array}\ee
Collecting \eqref{eq:thm.W1}, \eqref{eq:proof.lem.WI.T.a},
\eqref{eq:proof.lem.WI.T.2a}
and \eqref{eq:proof.lem.WI.T.3b}, the assertion \eqref{eq:lem.WI.S} follows.

To prove \eqref{eq:prob.S.k+1},
notice that the number of WI  balls of radius $L_k$
inside $\bcB^{(N)}(\Bx,L_{k+1})$
is bounded by the cardinality  $\sharp\bcB^{(N)}(\Bx,L_{k+1})$, and the probability for a WI ball
to be $(E,\delta ,m_N)$-S satisfies \eqref{eq:lem.WI.S},
so the last inequality in \eqref{eq:prob.S.k+1}) follows, again for $L_0$ large enough.
\qedhere


Now, given $k=0,1,\ldots$, consider the following probabilities:
$$
\bal
&\rP_k = \sup_{\Bu\in\bcZ} \pr{{\rm{ball}}\; \bcB^{(N)}(\Bu,L_k) \text{ is \EdmS}},
\\
&\rQ_{k+1} = 4 \sup_{\Bu\in\bcZ} \pr{\bcB^{(N)}(\Bu,L_{k+1}) \text{ is \EbR}},
\\
&\rS_{k+1} = \sup_{\Bx\in\bcZN} \pr{ \bcB^{N)}(\Bx,L_{k+1})
\text{ contains a WI \EdmS ball $\bcB^{(N)}(\Bu,L_k)$}}.
\eal
$$

Note that
\be\label{eq:N=1noWIb}
\hbox{for $N=1$, $\rS_{k+1}=0$ (there are no WI balls).}\ee

\begin{theorem}\label{thm:MSA.main.fixed}
Suppose that, for some given $\rB \ge 2$, $\delta\in (0,1)$,
$\kappa\in (0,\zeta)$  and $m^*,\nu^*\ge 1$, property {\rm\SSI{N,0}} holds true
with $L_0$ large enough. Then  {\rm\SSI{N,k}} holds true $\;\forall\;$ $k\ge 0$
with the same $L_0,\rB,\delta,\kappa,m^*$ and $\nu^*$.

\end{theorem}

\proof
It suffices to derive \SSI{N,k+1} from \SSI{N,k}, so assume the latter.
By virtue of Lemma \ref{lem:NR.NT.implies.NS}, if a ball $\bcB^{(N)}(\Bx ,L_{k+1})$ is \EdS,
then it is either \EbR (with probability $\le \quart Q_{k+1}$) or
$(E,\nu,K)$-bad.

By Eqn \eqref{eq:prob.S.k+1},
the probability of having at least one WI
\EmS ball $\bcB^{(N)}(\Bu,L_k)\subset\bcB^{(N)}(\Bx,L_{k+1})$ obeys:
$\rS_{k+1}\le \frac{1}{4 }\, \eu^{ - \mnu_N L_{k+1}^{\kappa}}$.

Note that this is the only point where the inductive hypothesis \SSI{N-1,k} is actually required,
for $N\ge 2$, while $\rS_{k+1}=0$ for $N=1$, because of \eqref{eq:N=1noWIb}.

Therefore, it remains to assess the probability of having a collection of  at least $K$ balls of radius $L_k$
inside $\bcB^{(N)}(\Bu,L_{k+1})$ which are SI, \EdS and pairwise $8NL$-distant.
The number of such collections is $\le C_\cZ^{KN} L_{k+1}^{KNd}$, thus, owing to Lemma \ref{lem:SI.full.sep}, we
have\footnote{A statement like Lemma \ref{lem:SI.full.sep} is not required for $N=1$, if the random
potential is IID, since the disjoint $1$-particle balls give rise to independent Hamiltonians.}
$$
\rP_{k+1} \le \half C_\cZ^{2N} L_{k+1}^{KNd} \rP_k^{K+1} + \rS_{k+1} + \frac{1}{4} \rQ_{k+1}.
$$
By Theorem \ref{thm:W1},
$\rQ_{k+1} \le C_\rW\, L_k^{(N+1)d}\eu^{-L_{k+1}^{\beta}}$, $ \beta > \kappa$,
so with $L_0$ large enough,
$Q_k \le \quart \eu^{-\mnu_N L_k^\kappa}$ for any $k\ge 0$.

Finally,
\be\label{verP}
\rP_{k+1} \le \half C_\cZ^{KN} L_{k+1}^{ KNd } \rP_k^{K+1}
 + S_{k+1} + \quart Q_{k+1}\ee
with $S_{k+1} + \quart Q_{k+1}\le\quart \eu^{-\mnu_N L_k^\kappa} + \quart \eu^{-\mnu_N L_k^\kappa}\le
\half \eu^{-\mnu_N L_k^\kappa}$.
The assertion of the theorem will follow if we show that, for $L_0$ large enough,
$\rP_{k+1} \le\eu^{-\mnu_N L_k^\kappa}$, i.e.,
$C_\cZ^{KN} L_{k+1}^{ KNd } \rP_k^{K+1}\le\eu^{-\mnu_N L_k^\kappa}$. The last fact can be
verified by using inequalities \eqref{eq:table1}.
This completes the proof of Theorem \ref{thm:MSA.main.fixed}.
\qedhere

Theorem \ref{thm:MSA.main.fixed} allows us to complete  the MPMSA inductive scheme.

In fact, owing to Theorem \ref{thm:L0.large.g}, $\,\forall\,$ given $L_0,\rB,\delta,\kappa,m^*$ and
$\nu^*$, property \SSI{N,0} holds true for sufficiently large $|g|$ and all $N=1,...,N^*$.
The scale induction step $k \rightsquigarrow k+1$ is provided by Lemmas \ref{lem:NR.NT.implies.NS}
and \ref{lem:WITRONS.subexp} and Theorems \ref{cor:prob.WI.S.large.g.DR} and
\ref{thm:MSA.main.fixed}. By induction in $k$, this proves \SSI{N,k}
for $1< N\le N^*$ and all $k\ge 0$, provided that \SSI{N-1,k} is proved for all $k\ge 0$.
The base of induction in $N$, is obtained in a similar (in fact, simpler) manner.

\subsection{From fixed to variable energy estimates} 
\label{ssec:FEMSA.to.VEMSA}

The core of the technical argument in this section is the two-volume EVC bound
(Theorem \ref{thm:W2.Vone}).

Given a positive integer $L$ and $\Bu\in\bcZ$, define the quantity  $\rBF_{\Bu}(E)=\rBF^{(N)}_{\Bu,L}(E)$:
\be\label{eq:def.Mxy}
\rBF_{\Bu}(E)= C_\cZ^{2N} L^{Nd} \max_{\Bz\in\pt^- \bcB (\Bx,L)}
\big| G^{(N)}_{\bcB(\Bu,L)}(\Bu,\Bz;E) \big|\,.
\ee
For brevity, we denote by $\Sigma_\Bu$ the spectrum of the operator $\BH^{(N)}_{\bcB (\Bu,L)}$.
Owing to assumption \Vone, the norms of the operators $\BH^{(N)}_{\bcB (\Bu,L)}$
are uniformly bounded, and so are their spectra. Therefore, the spectral analysis of these
operators can be restricted, without loss of generality, to some finite interval $I\subset\D{R}$
independent of $k\ge 0$ and $N=1,... ,N^*$.

Theorem \ref{thm:SW.ETV} encapsulates a probabilistic estimate essentially going back to the work
\cite{ETV10}. Its general strategy (converting fixed-energy probabilistic bounds into
those on the measure of "resonant" energies, with the help of the Chebyshev inequality) was employed
earlier in \cite{MS85}, and -- in a different context -- in \cite{BK05}.
Here we follow closely the book \cite{CS13}.

\begin{theorem}[Cf. \cite{CS13}*{Theorems 2.5.1 and 4.3.11}]\label{thm:SW.ETV}
Let be given an integer $L\ge 1$,  balls $\bcB^{(N)}(L,\Bx)$, $\bcB^{(N)}(L,\By)$,
an interval $I\subset\DR$ of length $|I|<\infty$ and numbers
$a_L, b_L, c_L, q_L>0$ satisfying
\be\label{eq:cond.a.b.c}
b_L \le \min\{K^{-1} a_L c_L^2, \, c_L\}.
\ee
with $K=\max\,\left[\sharp\bcB^{(N)}(L,\Bx),\sharp\bcB^{(N)}(L,\Bx)\right]$.
Suppose in addition that
\be
\pr{ \rBF_\Bx(E) \ge a_L} \le q_L, \;\; \pr{ \rBF_\By(E) \ge a_L} \le q_L .
\ee
Assume also that for some $A,C,>0$ and $\theta\in (0,1]$, $\forall$ $\eps >0$
\be\label{eq:assumed.W.ETV}
{\mathbb P}\,\Big\{ \dist\big(\Sigma_\Bx, \Sigma_\By \big) \le \eps\Big\} \le CL^A \eps^\theta .
\ee
Then one has, with $A' = A+2Nd$ and some $C'<\infty$,
\be
\pr{\exists\, E\in I:\,  \min \big[\rBF_\Bx(E), \rBF_\By(E) \big] \ge a_L }
\le \frac{2|I| q_L}{b} + C' L^{A'} c_L^\theta .
\ee
\end{theorem}

To prove Theorem \ref{thm:SW.ETV}, we need a general result, Theorem \ref{thm:Thm.2.5.1.RQSI}.
A similar assertion had been proven for  balls in an integer lattice $\DZ^d$ (cf. Theorem 2.5.1 in
\cite{CS13}). A direct inspection shows that the assertion remains true for $\bcZ=\cZ^N$ (with metric
$\brho$), and any finite subset $\bcV\subset\bcZ$. Here we state it for a ball $\bcB (\Bu,L)$, only
to set up a framework for the proof of Theorem \ref{thm:SW.ETV}.

\btm[Cf. Theorem 2.5.1 in \cite{CS13}]\label{thm:Thm.2.5.1.RQSI}
Given a positive integer $L$ and a configuration $\Bu\in\bcZ$, take the ball $\bcB =\bcB^{(N)}(\Bu,L)$.
Set $K = \sharp\,\bcB$ and
consider a random operator $\B{H}_\bcB=\B{H}^{(N)}_\bcB(\om)$ of the form
\be\label{HamBSA}
\left(\BH_\bcB f\right)(\Bx )=\left(-\Delta_\bcB f\right)(\Bx)+W(\Bx;\om )f(\Bx),
\;\;\Bx\in\bcB ,\ee
where $(\Bx,\om )\mapsto W(\Bx;\om )\in\DR$ is a given random function. (No condition is
imposed upon the distribution of $W(\Bx;\om )$.)
Let $E_j=E_j(\om )$, $1\leq j\leq K$,
be the (random) eigenvalues of $\B{H}_\bcB$
listed in some measurable way. Take a bounded interval $I\subset\DR$ and let the numbers
$a_L, b_L, c_L, q_L>0$ satisfy
\be
b_L \le \min\{K^{-1} a_L c_L^2, \, c_L\}
\ee
and $\forall$ $E\in I$
\be
\pr{ \rBF_\Bu(E) \ge a_L } \le q_L
\ee
where $\rBF_\Bu(E)$ is as in Eqn \eqref{eq:def.Mxy}. Then there is an event $\cC$ of probability $\pr{\cC}\le |I| b_L^{-1} q_L$ such that
for all $\om\not\in\cC$
\be
T_\Bu(2a_L) := \left\{ E\in I:\; \rBF_\Bu(E) \ge 2a_L\right\} \subset \cup_{j=1}^{K} I_j,
\ee
where $I_j := (E_j-2c_L,E_j+2c_L)$.
\etm
\medskip

\proof[Proof of Theorem {\rm{\ref{thm:SW.ETV}}}]
Let $\cC_\Bx$ and $\cC_\By$ be the events introduced in Theorem
\ref{thm:Thm.2.5.1.RQSI} for the balls $\bcB^{(N)}(L,\Bx)$ and $\bcB^{(N)}(L,\By)$,
and let $\cC = \cC_\Bx \cup\cC_\By$. As in  Theorem \ref{thm:Thm.2.5.1.RQSI}, denote
by $T_\Bx(2a_L)$, $T_\By(2a_L)$ the energy sets related to
$\bcB^{(N)}(L,\Bx)$, $\bcB^{(N)}(L,\By)$, and introduce the event
$\cE = \{\om:\, T_\Bx(2a_L) \cap T_\By(2a_L) \ne \varnothing \}$.
Then we have
\be\label{eq:proof.ETV.first}\bal
&\pr{ \cE  }
\le \pr{\cC} + \pr{ \cE\cap \cC^\rc}
\le 2 b_L^{-1} q_L |I| + \pr{  \cE \cap \cC^\rc}.
\eal\ee
For any $\om\in\cC^\rc$, each of the sets $\cE_\Bx(2a)$, $\cE_\By(2a)$ is covered by
intervals of length $4c_L$ centered at the respective EVs of $\BH^{(N)}_{\bcB (L,\Bx)}$ and
$\BH^{(N)}_{\bcB (L,\By)}$.

Recall that we have \emph{assumed} the two-volume Wegner-type estimate
\eqref{eq:assumed.W.ETV} (with an exponent $\theta >0$) for the pair
$\bcB^{(N)}(L,\Bx)$, $\bcB^{(N)}(L,\By)$.
(\textbf{N.B.}
By Theorem \ref{thm:W2.Vone}, we actually have $\theta =2/3$ for the
$N$-particle Hamiltonians satisfying Assumption \Vone, but here we keep a more
general assumption
\eqref{eq:assumed.W.ETV} and do not specify the value of $\theta$.)

Thus, with $\Sigma_\Bx=\Sigma(\BH^{(N)}_{\bcB (\Bx,L)})$
and $\Sigma_\By=\Sigma(\BH^{(N)}_{\bcB (\By,L)})$,
\be\label{eq:proof.ETV.final}
\bal
\pr{  \cE \cap \cC^\rc} &\le
\pr{ \dist(\Sigma_\Bx, \Sigma_\By) \le 4c_L }
\\
& \le(\sharp\,\bcB (\Bx,L)) \cdot (\sharp\,\bcB (\Bx,L))\cdot CL^A (2\cdot 4c_L)^{\theta}
\le{\ov C}L^{2Nd+A} c_L^{\theta},
\eal\ee
for some constant $C'$ and with $A'=A+2Nd$, as required. Collecting
\eqref{eq:proof.ETV.first} and \eqref{eq:proof.ETV.final}, the assertion of Theorem \ref{thm:SW.ETV} follows.
\qedhere

We use Theorem \ref{thm:SW.ETV} in the proof of Theorem \ref{thm:2vol.VEMSA.ETV}.

\smallskip

\begin{theorem}\label{thm:2vol.VEMSA.ETV}
Assuming $L_0$ large enough, $\forall$ $k$ and a pair of
$3NL_k$-distant balls $\bcB^{(N)}(\Bx,L_k)$, $\bcB^{(N)}(\By,L_k)$,
the following bound holds true:
$$
\pr{\exists\, E\in \DR:\;
\min\{ \rBF_\Bx(E), \rBF_\By(E)\} > \eu^{- m_N L_k^{\delta} } }
\le
\eu^{ -\frac{1}{9}\mnu_N L_k^\kappa } .
$$
\end{theorem}

\proof
As was noted, the spectrum $\Sigma_\Bx$ is contained in a fixed bounded interval
$I\subset\DR$. Given $k\ge 0$, we have, owing to property \SSI{N,k}, that, with
$a = \eu^{- m_N L_k^{\delta}}$,
$$\pr{\rBF_\Bx(E) > a }, \pr{\rBF_\By(E) > a } \le \eu^{ - \mnu_N L_k^\kappa}.$$
The LHS of  inequality \eqref{eq:assumed.W.ETV} can be assessed by virtue of
Theorem \ref{thm:W2.Vone}:
$$\pr{ \dist(\Sigma_\Bx, \Sigma_\Bx) \le \eps }  \le CL^{(2N+1)d} \eps $$
where $C=C^{(2)}_V$ is a constant.
Next, we are going to use  Theorem \ref{thm:SW.ETV}, with $L=L_k$ and
\be\label{eq:choice.a.b.c.}\begin{array}{c}
a_L = \eu^{-\mnu_N L^\kappa /3},\;
b_L = \eu^{-2\mnu_N L^\kappa /3},\;
c_L = \eu^{-\mnu_N L^\kappa /8},\;q_L=\eu^{-\nu_NL^\kappa},\\
A=2N+1,C=C^{(2)}_V,\theta =1.\end{array}\ee
We  obtain that
$$\begin{array}{r}
\pr{\operatornamewithlimits{\sup}\limits_{E\in I} \;
\min\big[\rBF_\Bx(E), \rBF_\By(E)\big] > \eu^{- m_N L_k^{\delta} } }\qquad\qquad{}\\
\diy\le 2|I| \exp\,\left(-\frac{\mnu_N}{3}L_k^\kappa\right)
   + {\ov C}L^{(4N+1)d} \exp\,\left(-\frac{\mnu_N }{8} L_k^\kappa\right).\end{array}$$
For $L_0$ large enough, the RHS can be made $\le \eu^{ -\mnu_N L_k^\kappa/9} $.
The assertion of Theorem \ref{thm:2vol.VEMSA.ETV} now follows.
\qedhere

Theorem \ref{thm:2vol.VEMSA.ETV} is important for the proof of Theorem \ref{thm:GF.decay.Holder}.

\ssect{Strong dynamical localization}  
\label{sec:MSA.to.DL}

In this section we complete the proof of assertion {\textbf{(A)}}
of Theorem \ref{thm:Main.DL.Vone.Uone.strong.g}. The staple here is
Theorem \ref{lem:GK} presenting a general result, under the key
assumption \eqref{eq:cond.thm.GK}.


Given a finite subset $\bcV\subset\bcZ$, we deal with
a random Hamiltonian $\BH_\bcV=\BH^{(N)}_\bcV(\om)$ on $\bcV$:
\be\label{HamYSA}
\left(\BH_\bcV f\right)(\Bx )=\left(-\Delta_\bcV f\right)(\Bx)+W(\Bx;\om )f(\Bx),
\;\;\Bx\in\bcV.\ee
Here
$(\Bx,\om )\mapsto W(\Bx,\om )$ is a bounded real-valued random field on $\bcV$. (As in
Theorem \ref{thm:Thm.2.5.1.RQSI},  no assumption
is made about the distribution of $W(\Bx,\om )$.) At the same time, we consider Hamiltonians
$\BH^{(N)}_{\bcB (\Bu,L)}$ in balls $\bcB^{(N)}(\Bu,L)\subseteq\bcV$. (Treating them as restrictions of
operator $\BH_\bcV$ to $\ell_2\left(\bcB^{(N)}(\Bu,L)\right)\subseteq\ell_2(\bcV)$.) Like before,
$G^{(N)}_{\bcV}(\Bu,\Bv;E)$ stands for the (random) GF
$\left\langle{\mathbf 1}_{\Bu},\left(\BH^{(N)}_{\bcV} -E\BI\right)^{-1}{\mathbf 1}_{\Bv}\right\rangle$.
Finally, as in Eqn \eqref{eq:def.Mxy},
\be\label{eq:def.Mxyg}
\rBF_{\Bu}(E)= C_\cZ^{2N} L^{d} \max_{\Bz\in\pt^- \bcB (\Bu,L)}
\big| G_{\bcB(\Bu,L)}(\Bu,\Bz;E) \big|.\ee

Like before, denote by $\csB_1(\DR)$ the set of all Borel functions
$\phi:\DR\to\DC$ with $\|\phi\|_\infty \le 1$.

\begin{theorem}\label{lem:5.1}
{\rm{(Cf. Lemma 9 in \cite{C12a})}}
\label{lem:GK}
Assume that $\forall$ positive integer $L$,
the following bound holds true for any
pair of disjoint balls $\bcB (\Bx ,L), \bcB (\By,L)\subset\bcS$ and some positive functions $u, h$:
\be\label{eq:cond.thm.GK}
\pr{ \exists\, E\in \DR:\,
\min\left[ \rBF_\Bx(E), \rBF_\By(E) \right] > u(L)}\le h(L).\ee
Then for any finite connected $\bcV\subset\bcS$ such that
$\bcV\supset\bcB(\Bx,L) \cup \bcB(\By,L)$,
\be\label{eq:thm.MSA.to.DL.sup}
\esm{ \sup_{\phi\in\csB_1(\DR)}\;  \big|\langle\one_\Bx | \phi(\BH_\bcV) | \one_\By \rangle \big| }
\le 4 u(L) + h(L).
\ee
\end{theorem}
\proof

The proof repeats verbatim that of Lemma 9 in \cite{C12a}, except for
the quantity $u(L)$ replacing an explicit expression $\eu^{-mL}$.
\qedhere


\vskip 5 truemm

Recall that $\kappa < \delta$ (cf. \eqref{eq:table1}).
For a finite $\bcV$, Assertion {\textbf{(A)}} follows from

\btm\label{thm:GF.decay.Holder}
Given $M>0$, $\exists$ $g_*=g_*(M)\in (0,\infty )$ and $C_*=C_*(M)\in (0,\infty )$ such that, for $|g|\ge g_*(M)$,
and $1\leq N\leq N^*$, $\forall$ $\Bx,\By\in\bcZ$ and a finite
$\bcV\subset\bcZ$ with $\bcV\ni\Bx,\By$,
\be\label{eq:dynlocinset}
\Upsilon_{\Bx,\By} :=
\esm{ \sup_{\phi\in\csB_1(\DR)}\;  \big|\langle\one_\Bx | \phi(\BH_{\bcV}) | \one_\By \rangle \big| }
\le  C_* \eu^{- M (\brho_{\rS}(\Bx,\By))^{\kappa}}.
\ee
\etm

\proof
Without loss of generality, it suffices to prove the assertion for the pairs of points with
$\brho_{\rS}(\Bx,\By)> 3NL_0$. Indeed, the EFC correlator is always bounded by $1$, so
for pairs $\Bx,\By$ with $\brho_{\rS}(\Bx,\By)\le 3NL_0$ the bound in \eqref{eq:dynlocinset} can be attained
by taking a sufficiently large constant $C_*$.

Thus, fix points $\Bx,\By\in\bcZ$ with $\rho: =\brho_{\rS}(\Bx,\By)> 3NL_0$. There exists
$k$ such that $\rho\in(3NL_k, 3NL_{k+1}]$. Arguing as above, it suffices to consider
a finite $\bcV\subset\bcZ$ such that
$
\bcB^{(N)}(\Bx,L_k) \cup \bcB^{(N)}(\By,L_k) \subset \bcV
$.

By Theorem \ref{thm:2vol.VEMSA.ETV} combined with Theorem \ref{lem:GK}, we have
\be\label{eq:locdynset}
\Upsilon_{\Bx,\By}
\le 4 \eu^{-m_N L_k^{\delta}} + \eu^{-\nu_N L_k^{\kappa}/9}.\ee
Since
$\rho \le 3NL_{k+1} = 3NB L_k$, we have that $L_k \ge \rho/(3NB)$.
With $|g|$ large enough, the initial scale estimate  \SSI{N,0} in \eqref{eq:SSIsexp} is fulfilled with
$m_N \ge 2 B N M$, $\nu_N \ge 10 N B M$ (cf. Theorem \ref{thm:L0.large.g}). Thus,
with $L_0$ large enough,
$$
\Upsilon_{\Bx,\By}
\le 5 \eu^{-10 N B M\rho^{\kappa}/(9NB)} \le \eu^{- M\rho^{\kappa}}.
$$
This completes the proof of Theorem \ref{thm:GF.decay.Holder}.
\qedhere

The case of an infinite $\bcV$ requires an additional limiting procedure (making use of the Fatou lemma
applied to the EF correlators),  developed in
\cite{AENSS03}*{Sect.~2}. As the argument can be
repeated here without any significant change, we omit it from the paper.

\section{Exponential decay of eigenfunctions} 
\label{Sec:exp.decay.EF}

The aim of this section is to prove assertion {\textbf{(B)}} of Theorem \ref{thm:Main.DL.Vone.Uone.strong.g},
about an exponential decay of the EFs. This is achieved along a scheme developed in
\cite{CS08,CS09a,CS09b,CS13} and modified to include the case of a graph $\cZ\in\fG (d,C)$
and an infinite-range interaction potential $U$. (We also use the same terminology.) In short, the
exponential decay follows when the inductive MPMSA scheme is successfully completed keeping the
mass parameter $m>0$; cf. \eqref{eq:thm.Main.DL.01}. In this section a particular form of $m=m_N$ is
used:  see \eqref{eq:table2}.

Compared to the scheme used in section \ref{Sec:FEMSA.subexp}, the main distinction is that
here we adopt a super-exponential scaling scheme where
\be\label{expLk}
L_{k+1} = \left\lfloor L_k^\alpha \right\rfloor ;
\ee
with the exponent $\alpha$ satisfying the conditions \eqref{eq:table2};
it depends upon the value of $\zeta$ in in condition \Uone (cf. Eqn \eqref{eq:def.U1}).
(The smaller $\zeta>0$, the larger is $\alpha$.)

The property \SSI{N,k} will be replaced in this section by its counterpart, \SSIexp{N,k}, presented in Eqn \eqref{eq:SSIexp},
adapted to the exponential decay bounds.
The verification of \SSIexp{1,k} (a one-particle case) is done in a standard way.

The relations between various parameters involved in the MSA inductive scheme
under Eqn \eqref{expLk} are summarised as follows.

\renewcommand{\arraystretch}{1.7}
\be\label{eq:table2}\hbox{\begin{tabular}{|l|l|}
  \hline
  $\tau > \max\left[\zeta^{-1}, 1 + \diy\frac{\ln (3N)}{\ln L_0}\right]$ & $0<\beta <\zeta\wedge 1\begin{matrix}\;\\
  \;\end{matrix}$
  \\
  \hline
  $\diy\max\left(\tau,\frac{3}{2}\right)< \alpha< \frac{7}{8\beta}$ &  $\begin{array}{l}\hbox{
  $L_0$ large enough, depending}\\
\hbox{\;\;\; upon $\alpha ,\beta ,\tau, K,m^*,\nu^*$} \end{array}$
\\
  \hline
  $m_N = m^*\,\big(1 + 4L_0^{-\delta+\beta}\big)^{N^*-N+1} $ & $m^*\ge 1$
\\
  \hline
  $P(N) = P^*\, (2\alpha)^{N^*-N+1}$ & $P^* > 4 N^* d\alpha $
\\
  \hline \end{tabular}}\ee

Like before, we assume the conditions   \eqref{eq:table2} throughout the whole section.

\ssect{The analytic step: scaling the GFs}\label{ssec:analyticstep} 


We continue our analysis of GFs of $\BH^{(N)}_{\bcB (\Bu,L)}$. The following definitions are modifications of Definitions \ref{def:NR.NS}
and  \ref{def:K.good.subexp}.

\bde\label{def:NS.exp}
Given $E\in\DR$ and $m^*\ge 1$, an $N$-particle ball $\bcB=\bcB^{(N)}(\Bu ,L)$ is called
$(E,m_N)$-nonsingular ({\rm{\EmNS}}), if $\;\forall$ $\By\in\pt^- \bcB$,
\be\label{eq:def.NSexp}
C_\cZ^{2N} L^{Nd}\cdot
\left|G^{(N)}_{\bcB}(\Bx,\By;E)\right| \le \eu^{-\gamma(m_N,L) L},
\ee
where
\be\label{eq:def.gamma.exp}
\gamma(m_N,L) := m_N(1+L^{-1/8}).
\ee
Otherwise, $\bcB$ is called $(E,m_N)$-singular ({\rm{\EmS}}).
\ede

\bde\label{def:K.good.exp}
Given $E\in\DR$, $\tau >0$, $m^*\ge 1$ and integers $k, K \ge 0$, we say that
a ball $\bcB^{(N)}(\Bu,L_{k+1})$ is $(E,m_N,K,\tau)$-good
($(E,m_N,K,\tau)$-{\rm G}) if  $\bcB^{(N)}(\Bu,L_{k+1})$ is $(E,\beta )$-{\rm{CNR}} (cf. Definition {\rm{\ref{def:CNR}}})
and contains no collection
of $\;\ge K+1$  pairwise $L_k^{\tau}$-distant
{\rm{\EmS}} balls of radius $L_k$. \ede

We will assume that $L_0$ is large enough so that $L_k^\tau > 8NL_k$
for all $k\ge 0$ (cf. lemma \ref{lem:SI.full.sep}).

These definitions are adopted throughout the current section.

A pre-requisite for the proof of the following statement is Appendix \ref{Sec:Ddecay}.

\ble\label{lem:CNR.and.K.EmS.implies.EmNS} {\rm{(Cf. Lemma \ref{lem:NR.NT.implies.NS}.)}}
If $\bcB^{(N)}(\Bu,L_{k+1})$ is $(E,m_N,K,\tau)$-{\rm G} and
$L_0$ is large enough, then $\bcB^{(N)}(\Bu,L_{k+1})$ is {\rm{\EmNS}}.
\ele

\proof
Set $\bcB =\bcB^{(N)}(\Bu,L_{k+1})$ and fix $\By\in\pt^{-}\bcB$. Set
$$ f_\By:\Bz \mapsto\left|G^{(N)}_{\bcB}(\Bz,\By;E)\right|.$$
The assumptions of Lemma \ref{lem:CNR.and.K.EmS.implies.EmNS} imply that $\exists$ a
(possibly empty) collection of
$K^\prime\le K$ balls $\bcB(\Bu_j,2L_k^\tau)\subset\bcB$ such that
any ball $\bcB (\Bv,L_k )\subset\bcB \setminus \cup_{j=1}^{K'} \bcB (\Bu_j,4NL_k)$ is \EmNS.
With $\bcL_r(\Bu)$ standing, as before, for a spherical layer
$\{\Bz\in\bcZ :\; \brho(\Bz ,\Bu )=r\}$, we set:
$$\Xi :=\left\{\Bx\in\bcB_{L_{k+1}-5L_k-1}(\Bu):\;
\bcL_{\rd(\Bu,\Bx)}(\Bu) \cap \cup_{j=1}^{K'} \bcB_{4NL_k}(\Bu_j) \ne \varnothing \right\}.$$
Then any ball $\bcB (\Bv,L_k )\subset\bcB$
with $\Bv \in\bcB\setminus \Xi$ is $(E,m_N)$-NS.
Also, define the function $\rR = \rR_{f,\Xi}$ as in Eqn \eqref{eq:rR.f.Xi}.

Owing to the assumptions of the lemma, the set $\Xi$ is covered by a family of annuli
with common center $\Bu$ and total width $\le 4KN L_k$.
By Lemma \ref{thm:GF.is.ell.q.Xi.dominated.subexp}
$f_\By$ is $(L_k,q,\Xi)$-dominated
in  $\bcB$. Here (cf. \eqref{eq:q.cond.nu.beta.delta} with $\delta =1$)
$$\bal
-\ln q &=  m_N (1 + L_k^{-1/8})L_k - L_{k+1}^{\beta} - \ln( C_\cZ^N L_{k+1}^{Nd})
\\
& \ge m_N L_k + \big(m_NL_k^{7/8} - 2 L_k^{\alpha\beta}  \big)
\ge L_k m_N \big(1 + {\textstyle\half} L_k^{-1/8} \big),\eal$$
where the last inequality follows from the form of $m_N$ in \eqref{eq:table2}.
By  virtue of Lemma
\ref{lem:rad.many.singular} (cf. Eqn \eqref{eq:bound.lem.rad.many.singular}),
$$
f_\By(\Bu) \le q^{\big(L_{k+1} - 2K(L_{k}^\tau + L_k)\big)/(1+L_k)} \rM(f, \bcB).
$$
One can see that $-\ln f_\By(\Bu)$ is
$$
\bal
\allowdisplaybreaks
&  \ge  -\ln \left\{
 \left( \eu^{-m_N(1 + L_k^{-1/8}/2)L_k} \right)^{(L_{k+1}- 4K L_k^\tau)/(1+L_k)}
   \eu^{ L_{k+1}^\beta}\right\}
\\
& = m_N \left(1+ L_{k}^{-1/8}/2 \right) L_{k+1}\;\cdot\;
\frac{ 1 - (4NK+1)L_{k+1}^{-1 + \frac{\tau}{\alpha}}  }{ 1 + L_{k}^{-1}} - L_{k+1}^{1/4}
\eal$$
which can be made
$$ \ge L_{k+1} m_N \left(1 + \frac{1}{4} L_{k}^{-1/8}
   - L_{k+1}^{-3/4} \right)
 \ge  \gamma(m_N, L_{k+1})L_{k+1} + \ln  \sharp\;(\pt\bcB),\;$$
assuming $L_0$ is large enough. This leads to the assertion of Lemma \ref{lem:CNR.and.K.EmS.implies.EmNS}.
\qedhere

\ssect{Localization in WI balls}\label{ssec:locWI} 

The main result of Section \ref{ssec:locWI} is Theorem \ref{cor:prob.WI.S.exp}.
We begin with an analog of Lemma \ref{lem:WITRONS.subexp}.

\begin{lemma}\label{lem:WITRONS.exp}
Fix $E\in\DR$ and consider a {\rm WI} ball $\bcB =\bcB^{(N)}(\Bu,L_k)$ with a
canonical factorization $\bcB = \bcB^\prime \times \bcB^{\prime\prime}$ and with reduced
Hamiltonians $\BH^\prime =\BH^{(N^\prime)}_{\bcB^\prime}$ and $\BH^{\prime\prime}=
\BH^{(N^{\prime\prime})}_{\bcB^{\prime\prime}}$ (cf. \eqref{BprimeBprimeprime}). Assume $\bcB$ is $(E,\beta)$-{\rm{NR}}.
Suppose in addition that $\forall\,\lam''\in\Sigma(\BH^{\prime\prime})$  the $N^\prime$-particle
ball $\bcB^\prime$  is $(E-\lam'', m_{N'})${\rm-NS} and $\forall\,\lam'\in\Sigma(\BH^\prime)$
the $N^{\prime\prime}$-particle ball
$\bcB^{\prime\prime}$  is $(E-\lam', m_{N''})${\rm-NS}.
Then ball $\bcB$ is $(E,\nu_N)${\rm-S}.
\end{lemma}
\proof See Section \ref{sec:proof.WITRONS.exp}. \qedhere
\medskip

Consider the following property (replacing \SSI{N,k}; cf. Eqn \eqref{eq:SSIsexp}).

\medskip\noindent
\SSIexp{N,k}: \qquad $\forall\, E\in\D{R}$,\; $1\leq n\leq N$ and $\Bu\in\bcZ^n$

\be\label{eq:SSIexp}
\pr{ \text{ball $\bcB^{(n)}(\Bu,L_k)$ is $(E,m_n)$-S } }
\le  L_{k}^{ -P(n) }.\ee

\begin{theorem}\label{cor:prob.WI.S.exp}
Suppose that property {\rm\SSIexp{N-1,k}} holds for some given $L_0,\alpha , \beta ,\tau$
and $m^*,P^*\ge 1$. Take $L_0$ large enough. Then $\forall$
$k\ge 0$ the following holds true.
Assume that  {\rm\SSIexp{N-1,k}} holds. Then $\forall$ $E$ and a WI ball $\bcB^{(N)} (\Bu,L_k)$,
\be\label{eq:lem.WI.S.exp}
\pr{\text{ $\bcB^{(N)} (\Bu,L_k)$  is $(E, m_N)${\rm-S} }}
\le  L_{k+1}^{ - \frac{3}{2} P(N) }.
\ee
Consequently, if $L_0$ is large enough then $\forall$ $E$ and ball $\bcB^{(N)}(\Bu ,L_{k+1})$,
\be\label{eq:prob.S.k+1.exp}
\begin{array}{r}
\pr{ \bcB^{(N)}(\Bu ,L_{k+1})
\text{ contains a {\rm WI} $(E, m_N)${\rm-S} ball of radius $L_{k}$}}
\qquad{}\\
\le C_\cZ^N L_{k+1}^{Nd} \cdot L_{k+1}^{ - \frac{3}{2} P(N) }
\le \diy\frac{1}{4 }  L_{k+1}^{- \frac{5}{4} P(N) }.\ea\ee
\end{theorem}

\proof First, we prove the bound in \eqref{eq:lem.WI.S.exp}. As in the propf of Theorem
\ref{cor:prob.WI.S.large.g.DR}, set $\bcB=\bcB^{(N)}(\Bu,L_k)$ and consider the
canonical factorization $\bcB=\bcB^\prime\times\bcB^{\prime\prime}$,
with reduced Hamiltonians $\BH^\prime$ and $\BH^{\prime\prime}$.
Given $E\in I$, introduce the event $\cS=\cS(E,N)$:
$$\cS= \{\om:\;  \bcB \text{ is {\rm WI} and $(E, m_N)${\rm-S}} \}.
$$
We have the following elementary inequality:
\be\label{eq:proof.lem.WI.T}
\bal
\pr{\cS } & < \pr{\text{ $\bcB$  is not  \FNR }}
\\ &+ \pr{\text{ $\bcB$  is  \FNR and $(E,m_N)$-S }}.\eal\ee
As earlier, the first term in the RHS of \eqref{eq:proof.lem.WI.T} is assessed in Theorem \ref{thm:FNR},
so we focus on the second summand.
Apply Lemma \ref{lem:WITRONS.subexp} and introduce events $\cS^\prime$ and
$\cS^{\prime\prime}$ by following the framework of Eqn \eqref{SprimeSprimeprime}
and \eqref{eq:estprob}. Then, with $m^\prime=m_{N^\prime}$,
$$\bal\pr{\cS^\prime }=\esm{ \pr{
 \exists\,\lam''\in\Sigma(\BH^{\prime\prime}):\, \bcB^\prime
   \text{ is  $(E-\lam'',m')$-S} \,\Big|\, \fF'' }}\,.\eal$$
By definition of the canonical decomposition, $
 \PPi \,\bcB^\prime \cap  \PPi \,\bcB^{\prime\prime} =\varnothing$,
and since the random field $V$ is IID, for any $E''\in \DR$, including
$E-\lam''$, the conditional probability  does not depend on the condition:
\be\label{eq:use.mixing.2}
\pr{\text{ $\bcB^\prime$  is $(E'', m')${\rm-S}} \,\big|\, \fF''}
\truc{\,=\,}{\rm{a.s.}}{}
\pr{\text{ $\bcB^\prime$  is $(E'', m')${\rm-S}} }.
\ee
On the other hand, by the assumed property \SSI{N-1,k},
\be\label{eq:proj.S.prob.2}
\pr{ \text{  $\bcB^\prime$  is $(E'', m')${\rm-S}} }
\le
C_\cZ^{-2N} L_k^{ - P(N-1)} = C_\cZ^{-2N} L_k^{ - 2 P(N)}.
\ee
Therefore, in analogy with \eqref{eq:estprob}, we obtain that
\be\label{eq:proof.lem.WI.T.2b}
\pr{\cS' } \le \sharp\;\bcB^{\prime\prime} \, \sup_{E''\in\DR}
\pr{ \text{$\bcB^\prime$  is $(E'', m_N)${\rm-S}} }
\le C_\cZ^N  L_k^{Nd}\,L_k^{ - P(N-1) };\ee
after the substitution $P(N-1)=2\alpha P(N)$ (cf. \eqref{eq:table2}),
the RHS can be made $\leq C_\cZ^N \half L_{k+1}^{ - \frac{3}{2}P(N) }$,
provided $P(N) > 2 Nd\alpha^{-1}$.
The latter inequality follows from the bound in \eqref{eq:table2}.

Summarising this calculation, we obtain
\be\label{eq:proof.lem.WI.T.2c}
\pr{ \text{$\exists\,\lam''\in\Sigma''$\,:\;
$\bcB^\prime$  is $(E - \lam'', m')${\rm-S}} }
\le \half  L_{k+1}^{ - \frac{3}{2}P(N) }.
\ee
\noindent
Similarly, with $ m'' = m_{N''}$,
\be\label{eq:proof.lem.WI.T.3c}
\pr{\text{ $\exists\,\lam'\in\Sigma'$\,:\; $\bcB''$  is $(E-\lam', m'')${\rm-S}}}
\le  \half  L_{k+1}^{ - \frac{3}{2}P(N) }.
\ee
Collecting \eqref{eq:thm.W1}, \eqref{eq:proof.lem.WI.T},
\eqref{eq:proof.lem.WI.T.2c}, \eqref{eq:proof.lem.WI.T.2a} and
\eqref{eq:proof.lem.WI.T.3c}, the assertion \eqref{eq:lem.WI.S.exp} follows.

To prove \eqref{eq:prob.S.k+1.exp},
notice that the number of WI  balls of radius $L_k$
inside $\bcB^{(N)}(\Bx,L_{k+1})$ is bounded by the cardinality  $\sharp\bcB^{(N)}(\Bx,L_{k+1})$,
and the probability that a given WI ball
is $(E,m_N)$-S satisfies \eqref{eq:lem.WI.S.exp}.
Therefore, the probability in the LHS of \eqref{eq:prob.S.k+1.exp} is upper-bounded,
for $L_0$ large enough, by
$$
C_\cZ^N L_{k+1}^{Nd} L_{k+1}^{- \frac{3}{2}P(N) }
=  L_{k+1}^{- \frac{5}{4}P(N) }  \cdot C_\cZ^N L_{k+1}^{- \frac{1}{4}P(N) + Nd}
\le \frac{1}{4} L_{k+1}^{- \frac{5}{4}P(N) }
$$
 since $P(N) \ge P(N^*) > 4Nd$, by virtue of \eqref{eq:table2}.
\qedhere

\ssect{The probabilistic scaling step}
\label{ssec:scale.ind.exp}

As in Section \ref{ssec:GreeninWI}, we introduce probabilities $\rP_k$, $\rQ_{k+1}$ and  $\rS_{k+1}$.
The following statement is a direct analog of Lemma \ref{lem:NR.NT.implies.NS} for the scaling scheme
\eqref{expLk}.

\ble\label{lem:NR.NT.implies.NS.exp}
Let us be given a positive integer $L_0$ and values $\alpha >0$, $\tau =1$, $\beta\in(0,1)$
and $m^*,P^* \ge 1$.
If a ball $\bcB^{(N)}(\Bu,L_{k+1})$ is  $(E,m_N,K,\tau)$-{\rm G} where $K =1$,
then it is $(E,m_N)$-{\rm NS}.
The assertion remains valid if the condition {\rm\EbNR}
(figuring in the definition of the $(E,\beta )$-{\rm{CNR}} property) is replaced by a weaker assumption:
\be\label{eq:lem.NR.NT.implies.NS.2}
\dist\left( \Sigma\left(\BH^{(N)}_{\bcB(\Bu,L_{k+1})}\right), E\right) \ge \eu^{-L^\beta}.
\ee
\ele

Lemma \ref{lem:NR.NT.implies.NS.exp} is a particular (and simpler) case of Lemma 4.2
in Ref.~\cite{DK89}. The latter has been adapted since then to various models and became a
common place. Hence, we omit its proof (it
is  similar to that of Lemma \ref{lem:NR.NT.implies.NS}).

Theorem \ref{thm:MSA.exp.fixed} is an analog of Theorem \ref{thm:MSA.main.fixed}.

\begin{theorem}\label{thm:MSA.exp.fixed}
Suppose that, for some given $\alpha >1$, $\tau$ as in \eqref{eq:table2}, $\beta\in (0,1)$,
and $m^*,P^*\ge 1$, property  {\rm\SSIexp{N,0}} is satisfied
with $L_0$ large enough. Then  {\rm\SSIexp{N,k}} holds true $\;\forall\;$ $k\ge 0$
with the same $L_0,\beta,\tau,m^*$ and $P^*$.
\end{theorem}

\proof
It suffices to derive \SSIexp{N,k+1} from \SSIexp{N,k}, so assume the latter.
By virtue of Lemma \ref{lem:NR.NT.implies.NS}, if a ball $\bcB (\Bu,L_{k+1})$ is $(E,\beta )$-S,
then
\par\noindent
$\bullet$
either $\bcB (\Bu,L_{k+1})$ is \EbR (with probability $\le \quart Q_{k+1}$),
\par\noindent
$\bullet$
or $\bcB (\Bu,L_{k+1})$ contains at least one WI $(E,m_N)$-{\rm NS} ball of radius $L_k$,
\par\noindent
$\bullet$
or $\bcB (\Bu,L_{k+1})$ contains at least two SI and $(E,m_N)$-{\rm S} balls of radius $L_k$.

By Eqn \eqref{eq:prob.S.k+1.exp},
the probability to have at least one WI
$(E,m_N)$-{\rm S} ball of radius $L_k$ inside $\bcB (\Bu,L_{k+1})$
obeys $\rS_{k+1}\le \frac{1}{4}\, L_{k+1}^{-P(N)}$.

Therefore, it remains to assess the probability to have a collection of at least two SI and
$(E,1 ,m_N)$-S balls of radius $L_k$ inside $\bcB (\Bu,L_{k+1})$.
The number of such collections is $\le C_\cZ^{2 N} L_{k+1}^{(2Nd}$, thus
$$
\rP_{k+1} \le \half C_\cZ^{2N} L_{k+1}^{2Nd} \rP_k^{2} + \rS_{k+1} + \rQ_{k+1}.
$$
By Theorem \ref{thm:W1},
$\rQ_{k+1} \le C_\rW\, L_k^{(N+1)d}\eu^{-L_{k+1}^{\beta}}$ where $ \beta > 0$ ,
thus for $L_0$ large enough,
$\diy Q_k \le \quart L_{k+1}^{-P(N)}$ for any $k\ge 0$.
Thus we can write
\be
\rP_{k+1} \le \half C_\cZ^{2N} L_{k+1}^{ 2Nd } \rP_k^{2}
 + \quart L_{k+1}^{-P(N)} + \quart L_{k+1}^{-P(N)},\ee
and the RHS can be made $< L_{k+1}^{-P(N)}$,
whenever $P(N) > 4 Nd$ and $L_0$ is large enough.
Again, the condition $P(N) > 4 Nd$ follows from \eqref{eq:table2}.
\qedhere

\subsection{Conclusion: exponential decay of eigenfunctions}

In this section, as before, the condition \Vone, as well as the property \RCM stemming from it
(cf. Theorem \ref{thm:RCM}), is always assumed, so we do not repeat it in the formulations of theorems
\ref{thm:2vol.3NL.balls} and \ref{thm:2vol.VEMSA.NonETV.exp}.

Recall that under the assumption \Vone, the spectrum of the Hamiltonian $\BH(\om)$, as well as the spectra
of its restrictions to arbitrary finite balls, is a.s. bounded by a value $O(|g|, N, d)$, so we can restrict our
analysis to a compact energy interval $I^*_g = I^*_g(N,d)\subset\DR$ of length  $|I^*_g|$.
Below we assume that such an interval is fixed.

An analog of Theorem \ref{thm:SW.ETV} is the following

\btm\label{thm:2vol.3NL.balls}
Suppose we are given two $3N L$-distant balls $\bcB_L(\Bx)$, $\bcB_L(\By)$ and numbers $a_L, q_L>0$ such  that for any $E\in\DR$
$$\max\Big[ \pr{ \rBF_\Bx(E) > a_L }, \pr{ \rBF_\By(E)  > a_L } \Big] \le q_L.$$
Then for any $b>0$, one has
\be\label{eq:thm.2vol.3NL.balls}
\pr{\exists\, E\in I^*_g:\,  \min(\rBF_\Bx(E), \rBF_\By(E)) \ge a_L } \le 2|I^*_g| b^{-1} q_L + \htil_L(4b),
\ee
where
\be
\htil(s) = C{\ov K}^2 L^A s^B + C' L^{A'} s^{B'},
\;\;{\ov K} = \max\{\sharp\,\bcB_L(\Bx), \sharp\,\bcB_L(\By)\}.
\ee
\etm

The reason why we need a separate bound \eqref{eq:thm.2vol.3NL.balls} is that the derivation of the variable-energy
estimates based on Theorem \ref{thm:SW.ETV} gives rise to exponential decay of eigenfunctions only if the probabilistic bounds
obtained in the fixed-energy analysis in the balls of size $L$ are also exponential in $L$; this can be seen in the
condition \eqref{eq:cond.a.b.c}.

In the proof given below, we will use the following auxiliary result.

\begin{theorem}[Cf. \cite{C14}*{Theorem 4}]\label{thm:2vol.VEMSA.NonETV.exp}
Suppose a ball $\bcB_L(\Bx)$ and numbers $a_L, q_L>0$ are such  that for all $E\in\DR$
\be\label{eq:prob.F.ge.a.q}
\pr{ \rBF_\Bx(E) > a_L } \le q_L.
\ee
Set $K =\sharp\,\bcB_L(\Bx)$. Then the following properties {\rm(A)}, {\rm(B)} hold true:

{\rm(A)} For any $b > q_L$ there exists an event $\cS_b$ with $\pr{\cS_b}\le b^{-1}q_L$
and such that for any $\om\not\in\cS_b$, the set of energies
$$
\cE_\Bx(a_L)  = \cE_\Bx(a_L;\om)  := \{ \rBF_\Bx(E) \ge a_L\}
$$
is covered by $K' < 3K$ intervals
$J_i = [E_i^-, E_i^+]$, of total length $\sum_i |J_i|\le b$.

{\rm(B)} Consider the parametric operator family $\BA(t) = \BH_\bcB + t\one$, $t\in\DR$.
The endpoints $E_i^{\pm}(t)$ for the operators $\BA(t)$ (replacing $\BH_{\bcB_L(\Bx))}$)
have the form
$$
E_i^{\pm}(t) = E_i^{\pm} + t, \;\; t\in\DR.
$$

\end{theorem}

{\it Proof of Theorem} \ref{thm:2vol.VEMSA.NonETV.exp}. (A)
Set for brevity $\bcB = \bcB_L(\Bx)$.
We have that
$$\rBF_\Bx = \max_{\By\in \pt^-\bcB} |\rBF_{\Bx,\By}|,\;\hbox{where}\;\rBF_{\Bx,\By}:=G_\bcB(\Bx,\By;E).$$
Fix $\By$ and consider $\rBF_{\Bx,\By}$ as  a rational function
\be\label{eq:GF.rational}
\rBF_{\Bx,\By}: E \mapsto \sum_{k=1}^K \frac{ c_k }{ E_k - E}
=  \sum_{k=1}^K \frac{ \langle \one_\Bx | \psi_k \rangle\,  \langle \psi_k |\one_\By \rangle }{ E_k - E}.
\ee
Its derivative is a ratio of two polynomials:
$$\frac{\rd}{\rd E}\rBF_{\Bx,\By}(E) = -\sum_k c_k(E_k - E)^{-2} =: \csP(E)/\csQ(E),$$
with $\deg \csP \le 2K - 2$. Hence, it has
$\le 2K-2$ zeros and $\le K$ poles, so $\rBF_{\Bx,\By}$ has $< 3K$ intervals of monotonicity.
Then the total number of monotonicity intervals for all functions $\rBF_{\Bx,\By}$ is upper-bounded by
$(\sharp\,\pt^- \bcB_{L}(\Bu)) \, \cdot 3K \le 3K^2$. Admitting the value $+\infty$ for the functions
$|\rBF_{\Bx,\By}|$, we can write
$$
\cup_{\By} \{E:\, |\rBF_{\Bx,\By}(E)| \ge a\} = \cup_{k=1}^K J_k = [E^-_k, E^+_k] \subset I^*_g .
$$
Let $\cS_{b,\Bx} = \{\om:\, \mes \{E\in I^*_g :\, \rBF_\Bx(E) \ge a \} \ge b\}$. By the Chebychev inequality
combined with the Fubini theorem, we have
\be\label{eq:Fubini}
\bal
\pr{ \cS_{b,\Bx} } & \le b^{-1} \esm{ \cS_{b,\Bx}} = b^{-1}\esm{ \int_{I^*_g} \one_{\{ \rBF_\Bx(E)\ge a \} } \, dE }
\\
& = b^{-1}\int_{I^*_g} \esm{\one_{\{\rBF_\Bx(E)\ge a \} }  }\, dE
   = b^{-1} \int_{I^*_g} \pr{\rBF_\Bx(E)\ge a }  \, dE
\\
& \le b^{-1} |I^*_g| q_L.
\eal
\ee
So, for all $\om\not\in\cS_{b,\Bx}$,
$\sum_k |J_k|\le \mes \{E\in I^*_g :\, \rBF_\Bx \ge a \} \le b$. This yields property (A).

\vskip1mm

(B) The operators $\BA(t) $ share common eigenvectors;
the latter determine the coefficients $c_k$ in \eqref{eq:GF.rational}, so we can choose the eigenfunctions
$\psi_k(t)$ constant in $t$ and obtain $c_k(t) \equiv c_k(0)$. The eigenvalues of $A(t)$ have the form
$E_k(t) = E_k + t$. Therefore, $\rBF_{\Bx,\By}(E;t) = \rBF_{\Bx,\By}(E-t;0)$, and
$J_k(t) = [E_k^- + t, E^+_k + t]$.
\qedhere

\proof[Proof of Theorem \ref{thm:2vol.3NL.balls}]
Fix $b>0$ and let $\cS_{b,\Bz} = \{\om: \mes\{E:\, \rBF_\Bz(E)\ge a\} \ge b\}$
for $\Bz\in\{\Bx,\By\}$, $\cS_b = \cS_{b,\Bx}\cup\cS_{b,\By}$.
Let $\cS$ be the event figuring in the LHS of \eqref{eq:thm.2vol.3NL.balls}. Using the bounds of the form \eqref{eq:Fubini}
on $\pr{\cS_{b,\Bx}}$ and $\pr{\cS_{b,\By}}$, we have
$$
\pr{ \cS } \le \pr{\cS_b} + \pr{ \cS \cap \cS_b^\rc} \le 2|I^*_g| b^{-1} q_L + \pr{ \cS \cap \cS_b^\rc}.
$$
It remains to asses $\pr{ \cS \cap \cS_b^\rc}$.

By Lemma \ref{lem:dist.WS}, the $3NL$-distant balls $\bcB_L(\Bx)$, $\bcB_L(\By)$ are weakly $\cB$-separated for some
$\cB\subset\bcZ$.
Consider the random variables $\xi = \xi_\cB = \lr{V(\cdot;\om)}_\cB$, $\eta_z(\om) = V(z;\om) -\xi(\om)$,
$z\in \cB$, and let $\fF_\cB$ be the sigma-algebra generated by $\{\eta_z, z\in \cB; V(u;\cdot), u\not\in \cB\}$.
Introduce the continuity modulus $\fs_\xi(\cdot|\fF_\cB)$ of the conditional probability distribution function
$F_\xi(t|\fF_\cB) = \pr{\xi \le t|\fF_\cB}$; it satisfies the condition \RCM with some $C',A',B',C'',A'',B''$.
The representation $V(z;\om) = \xi(\om) + \eta_z(\om)$ for $z\in \cB$ implies
$\BH_{\bcB} = n_1\xi(\om) + \BA(\om)$, where $\BA$ is $\fF_\cB$-measurable.

For any $\om\not\in\cS_b$, the energies $E$ where $\rBF_\Bx(E)\ge a$ are covered by a union of intervals
$J_{\Bx,i}$ with $|J_{Bx,i}|=:\eps_{\Bx,i}$, $\sum_i \eps_{\Bx,i} \le 2b$.
By  assertion (B) of Theorem \ref{thm:2vol.VEMSA.NonETV.exp}, we have
$$
J_{\Bx,i}(\om) = [ \lam_{\Bx,i}^- + n_1\xi(\om), \lam_{\Bx,i}^+ + n_1 \xi(\om) ],
$$
where $\lam_{\Bx,i}^\pm$ are $\fF_\cB$-measurable.

Similarly, introduce the intervals $J_{\By,j}$ with $|J_{By,j}|=:\eps_{\By,j}$, $\sum_i \eps_{\By,j} \le 2b$, and
$$
J_{\By,j}(\om) = [ \lam_{\By,j}^- + n_2\xi(\om), \lam_{\By,j}^+ + n_2 \xi(\om) ],, \;\; n_2 < n_1.
$$

We have
$$
\bal
& \{ \om:\, J_{\Bx,i}\cap J_{\By,j} \ne\varnothing \} \cap \cS_b^\rc \subset
\big\{ \om:\, |\lam_{\Bx,i} - \lam_{\By,j}| \le \eps_{\Bx,i} + \eps_{\By,j}\big\} \cap \cS_b^\rc
\\
&\quad
\subset \big\{ \om:\, |(n_1 - n_2)\xi  - \mu_{i,j}(\om)| \le 4b  \big\},
\eal
$$
with some $\fF_\cB$-measurable $ \mu_{i,j}$. Let $n = n_1 - n_2 \ge 1$ (recall that $\DZ\ni n_1-n_2>0$). By \RCM,
$$
\bal
\pr{ |(n_1 - n_2)\xi  - \mu_{i,j} | \le 4b  }  &\le
   \esm{  \pr{ |(n_1 - n_2)\xi  - \mu_{i,j} | \le 4b \,|\, \fF_\cB   } }
\\
& \le \pr{ \fs_\xi(4b|\fF_\cB) \ge C' L^{A'} (4b)^{B'} } + C' L^{A'} s^{B'}.
\eal
$$
Taking the sum over all $i$ and $j$, we obtain the asserted bound.
\qedhere

Setting $L=L_k$, $k\ge 0$, and
$$
a_{L_k} = \eu^{-m_N L_k}, \;\; q_{L_k} = L_k^{-P(N)}, \;\; b = L_k^{-P(N)/2},
$$
we come to the following  result, marking the end of the proof of our main theorem. Recall that
the strong dynamical localization bounds have already been established, and we only need to
prove exponential decay of the eigenfunctions.

\bco
For $k\ge 0$ and any pair of $3N L_k$-distant balls $\bcB_{L_k}(\Bx)$, $\bcB_{L_k}(\By)$ the following bound holds
true:
\be
\pr{\exists\, E\in\DR:\, \bcB_{L_k}(\Bx) \text{ and $\bcB_{L_k}(\By)$ are \EmS }}\le
C L_k^{-P(N)/2}.
\ee
Consequently, for $|g|$ large enough, with probability one, the operator $\BH^{(N)}_\bcV$, for any
$\bcV\subseteq\cZ^N$ such that $\bcV\supseteq\bcB_{L_k}(\Bx),\bcB_{L_k}(\By)$, has a
pure point spectrum, and its eigenfunctions obey {\rm{\eqref{eq:thm.Main.DL.01}}}.
\eco

\proof
The first assertion follows from Theorem \ref{thm:2vol.3NL.balls}. The second assertion is a well-known
result  going back to \cite{DK89}. In fact, the  proof of Lemma 3.1 from \cite{DK89} can be adapted to
pairs of balls $\bcB_{L_k}(\Bx),\bcB_{L_k}(\By)\subset\cZ^N$ at distance $\ge CL_k$, with a constant
$C\in(0,+\infty)$. The key fact is that structure of
the random potential (single- or multi-particle) is irrelevant to the proof of  \cite{DK89}*{Lemma 3.1}.
\qedhere

\appendices

\section{Proof of Lemmas \ref{lem:WITRONS.subexp} and \ref{lem:WITRONS.exp}}  
\label{Sec:proof.WITRONS.subexp}

\ssect{Proof of Lemma \ref{lem:WITRONS.subexp}}\label{ssect:AppendixA1}

\vskip2mm

\textbf{Step 1. Approximate decoupling.} In accordance with the canonical decomposition,
write $\Bu = (\Bu^\prime, \Bu^{\prime\prime})$ where $\Bu^\prime = \Bu_{\cJ}\in\bcZ^{N^\prime}$,
$\Bu^{\prime\prime}=\Bu_{\cJ^\rc}\in\bcZ^{N^{\prime\prime}}$. Let $\bcB =\bcB^\prime\times
\bcB^{\prime\prime}$ be the corresponding canonical factorization of the WI ball $\bcB =\bcB^{(N)}(\Bu,L_k)$
with $\bcB^\prime =\bcB^{(N^\prime )}(\Bu^\prime ,L_k)$, $\bcB^{\prime\prime}=
\bcB^{(N^{\prime\prime})}(\Bu^{\prime\prime},L_k)$. Cf. Eqns \eqref{eq:candcomp}--\eqref{eq:can.decomp}.

By Definition \ref{def:WI} and  Lemma \ref{lem:WI.decomp}, the graph distance between projected
configurations (in $\cZ$) satisfies $\rd\left( \Pi_\cJ \bcB ,  \Pi_{\cJ^\rc} \bcB \right) > L_k$,
yielding that $\forall$ $\Bx\in\bcB$,
$$
\rd (\Pi_{\cJ} \Bx, \Pi_{\cJ^\rc} \Bx ) > L_k.
$$

Consider representation \eqref{eq:Ham.decomposable}:
\be\label{eq:Hni}
\begin{array}{l}
\BH^{(N)}_{\bcB} =  \BH^{\rm{ni}}_{\bcB}+\BU_{\bcB^\prime ,\bcB^{\prime\prime}}\hbox{ where }\;
\BH^{\rm{ni}}_{\bcB}=\BH^{(N^\prime)}_{\bcB^\prime}\otimes \BI^{(N^{\prime\prime})}
+ \BI^{(N^\prime)} \otimes \, \BH^{(N^{\prime\prime})}_{\bcB^{\prime\prime}}.
\end{array}
\ee
(The superscript "$\rm{ni}$" stands for non-interacting.)
Here $\BU_{\bcB^\prime ,\bcB^{\prime\prime}}$ is the operator of multiplication by the function
\be\label{Umezhdu}
\begin{array}{l}
\Bx=(x_1,\ldots x_N)\in\bcB 
\mapsto \sum\limits_{1\leq i<j\leq N}{\mathbf 1}(i\in\cJ,
j\in\cJ^\rc)\,U(\rd (x_i,x_j)).
\end{array}
\ee
According to assumption \Uone, the norm of operator $\BU_{\bcB^\prime ,\bcB^{\prime\prime}}$ obeys
\be\label{eq:bound.interaction.dist.L.subexp}
\left\|\BU_{\bcB^\prime ,\bcB^{\prime\prime}}\right\|
 \le C (N^\prime \cdot N^{\prime\prime}) \eu^{-L_k^\zeta} \le CN^2 \eu^{-L_k^\zeta},
\ee
with $C=C_U$ as in \eqref{eq:def.U1}.

The eigenvalues of $\BH^{\rm{ni}}_{\bcB}$ are the sums $E_{a,b}=\lam_a+\mu_b$, where
$\lam_a$ form the spectrum $\Sigma\left(\BH^{(N^\prime)}_{\bcB^\prime}\right)$
and $\mu_b$ the spectrum $\Sigma\left(\BH^{(N^{\prime\prime})}_{\bcB^{\prime\prime}}\right)$.
The eigenvectors of $\BHBni$ can be chosen in the form
$\Bphi_a \otimes \Bpsi_b$
where $\{\Bphi_a\}$ are eigenvectors of $\BH^{(N^\prime)}_{\bcB^\prime}$ and
$\{\Bpsi_b\}$ of $\BH^{(N^{\prime\prime})}_{\bcB^{\prime\prime}}$.

\vskip2mm
\noindent
\textbf{Step 2. Nonresonance properties.}
Next we infer from the assumed \EbNR property of $\bcB$ (with regard to the
resolvent $\B{G}^{(N)}_\bcB(E)=(\b{H}^{(N)}_\bcB-E\B{I})^{-1}$) a similar (albeit a weaker) property for
the resolvent $\B{G}^{\rm{ni}}(E)=(\b{H}^{\rm{ni}}_\bcB-E\B{I})^{-1}$. By the min-max principle,
\be\label{eq:perturbed.nonres.subexp}
\begin{array}{r}
\dist\Big(\Sigma(\BHBni), E \Big)
\ge  \dist\left(\Sigma\left(\B{H}^{(N)}_\bcB\right), E \right) - \left\|\BU_{\bcB^\prime ,\bcB^{\prime\prime}}\right\|
\qquad{}\\
\ge
2\eu^{ -L_k^\beta } - C_U \eu^{ -4L_k^\zeta } \ge  \eu^{ -L_k^\beta },
\end{array}\ee
provided that $\beta<\zeta$ (which is one of conditions  \eqref{eq:table1})
and $L_0$ is large enough.

For each pair $(\lam_a,\mu_b)$, the non-resonance condition $|E - (\lam_a + \mu_b)|\ge \eu^{L_k^\beta}$
reads as $|(E -\lam_a) - \mu_b)|\ge \eu^{L_k^\beta}$ and also as
$|(E -\mu_b) - \lam_a)|\ge \eu^{L_k^\beta}$.
In terms of resolvents  $\B{G}^{(N)}_\bcB(E)$ and  $\BGBni(E)$ we then have:
\be\label{eq:norm.GB.GBni.subexp}
\left\| \B{G}^{(N)}_\bcB (E)\right\| \le \half \eu^{L_k^\beta} < \eu^{L_k^\beta}, \;\;
\| \BGBni(E) \| \le \eu^{L_k^\beta}.\ee

\vskip2mm
\noindent
\textbf{Step 3. Analytic perturbation estimates.}
We begin with analyzing the resolvent $\BGBni(E)$. Start with the identities for
the GF $\;G^{\rm{ni}}_\bcB (\Bu,\By; E)$:
\begin{eqnarray}
\label{eq:GF.decomposition.subexp}
G^{\rm{ni}}_\bcB(\Bu,\By; E) &=& \sum_{\lam_a} \sum_{\mu_b}
\frac{ \Bphi_a(\Bu') \Bphi_a(\By')\, \Bpsi_b(\Bu'') \Bpsi_b(\By'')\, }
{ (\lam_a + \mu_b) - E}
\\
\label{eq:a.subexp}
& =& \sum_{\lam_a}\Bphi_a(\Bu') \Bphi_a(\By') \, G^{(N^{\prime\prime})}_{\bcB^{\prime\prime}}(\Bu'',\By''; E-\lam_a)
\\
\label{eq:b.subexp}
& =& \sum_{\mu_b}  \Bpsi_b(\Bu'') \Bpsi_b(\By'') \, G^{(N^\prime )}_{\bcB^\prime}(\Bu',\By'; E-\mu_b).
\end{eqnarray}
By assumptions of the lemma,
\be\label{eq:projections.NS.subexp}
\bal
&\bullet
\text{$\forall$ $\mu_b\in\Sigma\left(\BH^{(N^{\prime\prime})}_{\bcB^{\prime\prime}} \, \right)$, the ball
$\bcB^\prime$ is $(\mu_b,\delta,\mnu_{N^\prime })$-NS,}
\\
&\bullet \text{$\forall$  $\lam_a\in\Sigma\left(\BH^{(N^\prime)}_{\bcB^\prime} \,\right)$,  the ball
$\bcB^{\prime\prime}$ is $(\lam_a,\delta, \mnu_{N^{\prime\prime}})$-NS.}
\eal
\ee
\noindent
For any $\By\in\pt^- \bcB$, either $\brho^{(N^\prime)}(\Bu',\By') = L_k$ or
$\brho^{(N^{\prime\prime})}(\Bu'',\By'') = L_k$. In the first case we infer from  \eqref{eq:b.subexp},
combined with $(\mu_b,\delta,\mnu_{n'})$-NS property of ball $\bcB^\prime$, that
\be\label{eq:Gni.1}
\big|G^{\rm{ni}}_\bcB (\Bu,\By; E) \big| \le\sharp\,\bcB^{\prime\prime}\,
\eu^{- m_{N^\prime} L_k^{\delta} + 2L_k^{\beta}}.\ee
Similarly, in the second case we obtain that
\be\label{eq:Gni.2}
\big|G^{\rm{ni}}_\bcB(\Bu,\By; E) \big| \le \sharp\,\bcB^\prime\,
\eu^{- m_{N^{\prime\prime}} L_k^{\delta} + 2L_k^{\beta}}.
\ee
In either case, the LHS is bounded by (cf. \eqref{eq:def.mN})
\be\label{eq:proof.WITRONS.factor.nu.N}
C_\cZ^N L_k^{Nd}\eu^{ -m_{N-1} L_k^{\delta} + 2L_k^{\beta} }
\le  \eu^{ -m_N L_k^{\delta} - L_0^\beta }\le {\textstyle \half} \eu^{ -m_N L_k^{\delta} },\ee
provided that $L_0$ is large enough.

Now, to assess $G^{(N)}_\bcB (\Bu,\By; E)$, we use the second resolvent equation
and write:
\be\label{eq:proof.WITRONS.step3.1}
\bal
\left\|\B{G}^{(N)}_\bcB (E) - \BGBni(E)\right\|  & \le  \| \BGBni(E)\| \,\left\|
\B{U}_{\bcB^\prime,\bcB^{\prime\prime}}\right\| \, \left\|\B{G}^{(N)}_\bcB(E)\right\|
\\
& \le C_U \eu^{2 L_k^\beta - L_k^{\zeta}} \le \eu^{- \frac{1}{2} L_k^{\zeta}}
\le \half \eu^{ -\mnu_N L_k^{\delta} },
\eal\ee
provided that $\beta < \delta < \zeta$ and $L_0$ large enough (this
holds in accordance with \eqref{eq:table1}).

Collecting  \eqref{eq:a.subexp}, \eqref{eq:b.subexp},
\eqref{eq:proof.WITRONS.factor.nu.N} and \eqref{eq:proof.WITRONS.step3.1}, we get
\be
\bal
 \max_{\By\in \pt^- \bcB} \left| G^{(N)}_\bcB (\Bu,\By; E)\right| &
\le\half \eu^{ -\mnu_N L_k^{\delta} }
\; + \,\half \eu^{ -\mnu_N L_k^{\delta} }
=\eu^{ -\mnu_N L_k^{\delta} } \, .
\eal\ee
We see that ball $\bcB$ is \mEdmNS{\delta,\mnu_N}.
\qedhere

\ssect{Proof of Lemma \ref{lem:WITRONS.exp}}  
\label{sec:proof.WITRONS.exp}

The line of the argument here follows, {\it mutatis mutandis}, that from the proof of Lemma
\ref{lem:WITRONS.subexp}.

\vskip2mm
\noindent
\textbf{Step 1. Approximate decoupling.} We start as in the previous section,
but  have to achieve an exponential bound upon the GFs. The
bound on the interaction \eqref{eq:bound.interaction.dist.L.subexp} is to be
modified accordingly:
\be\label{eq:bound.interaction.dist.L.exp}
\brho  \left( \Pi_\cJ \bcB, \Pi_{\cJ^{\rm c}}\bcB\right) \ge L_k^\tau\;\hbox{ and }\;
\left\|\BU_{\bcB^\prime ,\bcB^{\prime\prime}}\right\|\le C_UN^2 \eu^{-ML_k}.
\ee
Here $M$ can be chosen arbitrarily large, provided that $L_0$ is large enough. Specifically,
we require that $M \ge \max\{1, m_1\}$, hence $M \ge \max\{1, m_N\}$ for $1 \le N \le N^*$.
Cf. \eqref{eq:table2}.

The definitions of the operators $\BH^{\rm{ni}}_{\bcB}$ and $\BU_{\bcB^\prime ,\bcB^{\prime\prime}}$
(see \eqref{eq:Hni}  \eqref{Umezhdu}) remain in force.

\vskip2mm
\noindent
\textbf{Step 2. Nonresonance properties.}
A direct analog of \eqref{eq:perturbed.nonres.subexp} is
\be\label{eq:perturbed.nonres.exp}
\dist\left[\Sigma(\BHBni), E \right]\ge
2\eu^{ -L_k^\beta } - \eu^{ - M L_k } \ge  \eu^{ -L_k^\beta };\ee
it implies, as before, that
\be\label{eq:norm.GB.GBni.exp}
\left\|\B{G}^{(N)}_\bcB(E)\right\| \le \half \eu^{L_k^\beta} < \eu^{L_k^\beta}, \;\;
\| \BGBni(E) \| \le \eu^{L_k^\beta}.
\ee

\vskip2mm
\noindent
\textbf{Step 3. Analytic perturbation estimates.}
We can use identities \eqref{eq:GF.decomposition.subexp}--\eqref{eq:b.subexp} and
the (assumed) properties \eqref{eq:projections.NS.subexp}. Now,
the estimates \eqref{eq:Gni.1}--\eqref{eq:Gni.1} are to be modified as follows.

Given $\By\in\pt^- \bcB$, we again have two possibilities. (i)
$\brho^{(N^\prime )}(\Bu',\By') = L_k$, in which case we deduce from  \eqref{eq:b.subexp},
combined with $(\mu_b,m_{N^\prime})$-NS property of the ball $\bcB^\prime$, that
\be\label{eq:Gni.1.exp}
\big| G^{\rm{ni}}(\Bu,\By; E) \big| \le \sharp\,\bcB^{\prime\prime}\,
\eu^{- m_{n'} L_k + 2L_k^{\beta}}.
\ee
The other case is where (ii)  $\brho^{(N^{\prime\prime})}(\Bu'',\By'') = L_k$ -- then, similarly
to \eqref{eq:a.subexp}, we have that
\be\label{eq:Gni.2.exp}
\big|G_\bcB^{\rm{ni}}(\Bu,\By; E) \big| \le\sharp\,\bcB^\prime\,
\eu^{- m_{n''} L_k + 2L_k^{\beta}}.
\ee
In either case, with $C_\cZ^N L_k^{Nd} \le \eu^{L_k^\beta}$, the LHS of Eqn
\eqref{eq:a.subexp} is bounded by
\be\label{eq:proof.WITRONS.factor.nu.N.exp}
C_\cZ^N L_k^{Nd}\eu^{ -m_{N-1} L_k + 2L_k^{\beta} }
\le  \eu^{ - m_N L_k - L_0^{\beta} }
\le \half \eu^{ -m_N L_k } .\ee
Now, by virtue of the second resolvent identity we have
\be\label{eq:proof.WITRONS.step3.1.exp}
\bal
\left\|\B{G}^{(N)}_\bcB (E) - \BGBni(E)\right\|  & \le  \| \BGBni(E)\| \, \|\BUJ\| \,\left\|\B{G}^{(N)}_\bcB (E)\right\|
\le \half \eu^{ - m_N L_k } ,
\eal
\ee
since $M \ge 1$, $L_0 > 1$,  $\beta \le 1$.
Collecting  \eqref{eq:a.subexp}--\eqref{eq:b.subexp} and the bounds \eqref{eq:proof.WITRONS.factor.nu.N.exp}--\eqref{eq:proof.WITRONS.step3.1.exp}, we
obtain
$$
\bal
 \max_{\By\in \pt^- \bcB}
\left| G^{(N)}_\bcB(\Bu,\By; E) \right|
\le \half \eu^{ - m_N L_k }
\; + \,\eu^{- 2M L_k }
\le \eu^{ -m_N L_k } \, .
\eal
$$
Therefore, ball $\bcB$ is $(E,m_N)$-NS.
\qedhere

\section{Dominated decay of functions on $\bcZ$}  
\label{Sec:Ddecay}

In this section we establish Lemmas \ref{lem:rad.many.singular} and \ref{thm:GF.is.ell.q.Xi.dominated.subexp} applicable to arbitrary
locally finite, connected graphs, including $\bcZ=\cZ^N$, $N\ge 2$. These lemmas are related to
iterations of the GRI (see Eqn \eqref{eq:GRE}) and provide an
ingredient in the proof of  Lemma \ref{lem:NR.NT.implies.NS} and \ref{lem:CNR.and.K.EmS.implies.EmNS}.
The argument here stems from \cite{DK89}, Lemma 4.2; in the case where $\cZ=\D{Z}^d$,
it was presented in \cite{CS13}, Sect 2.6.

\bde 
\label{def:ell.q.Xi.dominated.f} {\rm{(Cf. Definition 2.6.1 in  \cite{CS13})}}
Let us be given a finite
subset $\bcV\subset\bcZ$, a non-negaitive function $f:\bcV\to [0,\infty )$, a number $q\in(0,1)$ and two
integers $L \ge \ell \ge 1$. Take an $N$-particle ball $\bcB (\Bu,L)\subset\bcV$.
\par\vskip1mm

{\rm{(1)}} A point $\Bx\in\bcB^{(N)}(\Bu,L-\ell )$ is called $(\ell,q)$-regular for the function $f$, if
\be
f(\Bx) \le q\, \rM(f, \bcB (\Bx,\ell+1)),
\ee
and $(\ell,q)$-singular, otherwise. Here and below, we set:
\be \rM(f,\bcW )=\sup\,\big[f(\By ):\;\By\in\bcW\big],\;\;\bcW\subseteq\bcV.
\ee
The set of all $(\ell,q)$-regular points $\Bx\in\bcB (\Bu,L)$ for $f$ is denoted by
$\csR_f(\Bu)=\csR_{f,q,\ell}(\Bu)$, and the set of all $(\ell,q)$-singular points by
$\csS_f(\Bu)=\csS_{f,q,\ell}(\Bu)$.
\par\vskip1mm

{\rm{(2)}}  A spherical layer
$$\bcL_r(\Bu)=\big\{\By\in\bcZ:\;\rd (\Bu,\By )=r\big\}$$
is called regular if $\bcL_r(\Bu )\subset \csR_f(\Bu)$.
\par\vskip1mm

{\rm{(3)}} For $\Bx\in\bcB (\Bu ,L-\ell )$,  set \def\ovr{{\overline r}}
$$
\ovr(\Bx) :=\begin{cases}
\min\big[r \ge \rd(\Bu,\Bx): \, \bcL_r(\Bu)\subset \csR_f(\Bu )= \varnothing ],\\
\qquad\quad\,\hbox{if a regular layer $\bcL_r(\Bu)$ exists, with $r\ge\rd (\Bu,\Bx)$,}\\
+\infty ,\quad\hbox{if no such layer $\bcL_r(\Bu )$ exists,}
\end{cases}$$
and
\be\label{eq:rR.f.Xi}
\rR_f(\Bx) (= \rR_{f,q,\ell}(\Bx)) =
\left\{
  \begin{array}{ll}
   \ovr(\Bx)+\ell  , & 
                             \ovr(\Bx)<+\infty , \\
    +\infty, & \hbox{otherwise.} \\
  \end{array}
\right.
\ee
\par\vskip1mm

{\rm{(4)}} Given a set $\Xi\subseteq\bcV$, function $f$ is called $(\ell,q,\Xi)$-dominated
in $\bcB (L,\Bu)$ if $\csS_f(\Bu)\subset\Xi$, and
  for any $\Bx\in\bcB (\Bu ,L-\ell)$ with $\rR_f(\Bx) <+\infty$, one has
\be\label{eq:bound.x.rR.finite}
f(\Bx) \le q \rM(f, \bcB (\Bu,\rR_f(\Bx))).
\ee

\ede

\ble[Cf. Theorem 2.6.1 in \cite{CS13}]\label{lem:rad.many.singular} 
Let function $f:\bcV\to\DR_+$
be $(\ell,q,\Xi)$-dominated in an $N$-particle ball $\bcB(\Bu ,L)$, where $L\ge \ell\ge 0$.
Assume that set $\Xi$ is covered by a union $\bcU$ of concentric annuli
$\bcB (\Bu ,b_j)\setminus \bcB (\Bu ,a_j - 1)$, with
$$
\textstyle
\rw(\bcU) := \sum_j (b_j - a_j +1)\le L - \ell .
$$
Set: $\rW =\rW (L,\ell, \bcU):= \diy\frac{L+1 - \rw(\bcU)}{\ell+1}$. Then
\be\label{eq:bound.lem.rad.many.singular}
f(\Bu) \le q^{\lfloor\rW\rfloor} \rM(f, \bcB (\Bu ,L+1))
\le q^{\rW} \rM(f, \bcB (\Bu ,L+1)).
\ee
\ele

The proof of Lemma \ref{lem:rad.many.singular} repeats {\it verbatim} that
of Theorem 2.6.1 in \cite{CS13}, and we omit from the paper.
\medskip

\ble[Cf. Theorem 2.6.2 in \cite{CS13}]\label{thm:GF.is.ell.q.Xi.dominated.subexp} 
Fix $0<\beta ,\delta \leq 1$, $m>0$ and $E\in\DR$.
Suppose that, for some integer $L\ge 1$ and $\Bu\in\cZ^N$,
the $N$-particle ball $\bcB (L,\Bu)$ is $(E,\beta )$-{\rm{CNR}}. Next, take a finite $\bcV\subset\cZ^N$ such
that $\bcV\supset\bcB (\Bu,L)$ and $\By\in\bcV\setminus\bcB (\Bu,L)$, and
consider the function
\be\label{functionf}
f = f_{\By ,\bcV}:\; \Bx\in\bcB (L,\Bu) \mapsto\left|G^{(N)}_{\bcV}(\Bx,\By; E)\right| .
\ee
Given $\ell =0,\ldots ,L-1$, let $\Xi = \Xi(E)\subset\bcB(\Bu ,L-\ell-1)$
be a (possibly empty) subset such that
any ball $\bcB (\Bx,\ell)\subset \bcB (\Bu,L-\ell-1) \setminus \Xi$ is \mEdmNS{\delta,m}.

If
\be\label{eq:cond.nu.beta.delta}
m \ell^\delta > 2L^\beta > L^\beta + \ln\left(C_\cZ L^D\right)
\ee
then $\forall$ $\By\in\pt^- \bcB (\Bu ,L)$, function $f$
is $(\ell,q,\Xi)$-dominated in $\bcB (\Bu ,L)$, with
\be\label{eq:q.cond.nu.beta.delta}
q = \eu^{-m' \ell^\delta}, \;\hbox{ where }\;
m' := m - 2 \ell^{-\delta} L^\beta  .\ee
\ele

\proof
First note that for any $\Bx\in\bcB (\Bu ,L-\ell) \setminus\Xi$ we have
$$
f(\Bx) \le \eu^{-m \ell^\delta} \, \rM\big(f, \bcB (\Bx ,\ell) \big),
$$
since ball $\bcB (\Bx ,\ell )$ must be \mEdmNS{\delta,m}, by definition of set $\Xi$.
Clearly, $\eu^{-m \ell^\delta}<q$, where $q$ is given by
\eqref{eq:q.cond.nu.beta.delta}.

Further,
define the function $\Bx \mapsto \rR_f(\Bx)$ in the same way as in \eqref{eq:rR.f.Xi}.
Suppose that $\Bx\in\Xi$ and $\rR_f(\Bx) < \infty$, i.e., the spherical layer
$\cL_{\rR_f(\Bx)}(\Bu)$ is regular, i.e., each point $\By\in\cL_{\rR_f(\Bx)}(\Bu)$ is regular.
Set for brevity $r^* = \rR_f(\Bx)$.
Applying the GRI \eqref{eq:GRE} to the ball $\bcB (\Bu, r^* -1)$, we get
$$\bal f(\Bx) &
\le C_\cZ ({r^*})^D \|\BG_{\bcB (\Bu, r^*-1)}(E)\|
\cdot \max_{\Bz\in \bcL_{r^*}(\Bu)} |G_{\bcB (\Bu,r^*)}(\Bz,\By;E)|
\\
&
\le C_\cZ L^{D} \eu^{L^\beta} \, \rM(f, \bcL_r(\Bu)).
\eal
$$
Next, applying the GRI to each ball $\bcB(\Bz,\ell)$ with
$\Bz\in \bcL_{r}(\Bu)$, we obtain
$$
\bal
f(\Bx) & \le C_\cZ L^{D} \eu^{L^\beta} \, \eu^{-m \ell^\delta}
\rM(f, \bcL_{r+\ell}(\Bu))
\le
\eu^{- m' \ell^\delta  }
\rM(f, \bcL_{r+\ell}(\Bu)),
\eal
$$
with $m'$ given by \eqref{eq:q.cond.nu.beta.delta},
provided that the condition \eqref{eq:cond.nu.beta.delta} is fulfilled.
Thus $f$ is indeed $(\ell,q,\Xi)$-dominated in $\bcB (\Bu,L)$, with
$q$ given by \eqref{eq:q.cond.nu.beta.delta}.
\qedhere
\medskip

Lemma \ref{thm:GF.is.ell.q.Xi.dominated.subexp} is used in the proof of
Lemmas \ref{lem:NR.NT.implies.NS} and \ref{lem:CNR.and.K.EmS.implies.EmNS}.


\section*{Acknowledgements} VC thanks the Gakushuin University of Tokyo, the Kyoto University
and the Research Institute for Mathematical Science (RIMS, Tokyo) for the warm hospitality in December 2013,
and Prof. F. Nakano, S. Kotani and N. Minami
for stimulating discussions of localization properties of disordered quantum systems.
YS thanks IME USP,
Brazil, for the warm hospitality during the academic year of 2013-4. The authors
thank Abel Klein, G\"{u}nter Stolz, Ivan Veseli\'c and Peter M\"{u}ller for fruitful discussions.



\begin{bibdiv}
\begin{biblist}



\bib{AENSS03}{article}{
      author={Aizenman, M.},
     author={Elgart, A.},
      author={Naboko, S.},
      author={Schenker, J.~H.},
      author={Stolz, G.},
       title={Moment analysis for localization in random schr\"odinger
    operators},
        date={2006},
     journal={Invent. Math.},
      volume={163},
       pages={343\ndash 413},
}

\bib{AW09b}{article}{
      author={Aizenman, M.},
      author={Warzel, S.},
       title={Complete dynamical localization in disordered quantum
  multi-particle systems},
        date={2009},
     journal={XVIth Int. Congress on Math. Phys., World Sci., 556-565},
      volume={{}},
}

\bib{AW09a}{article}{
      author={Aizenma{n}, M.},
      author={Warzel, S.},
       title={Localization bounds for multi-particle systems},
        date={2009},
     journal={Commun. Math. Phys.},
      volume={290},
       pages={903\ndash 934},
}



\bib{BAA06}{article}{
      author={Basko, D.M.},
      author={Aleiner, I.L.},
      author={Altshuler, B.L.},
       title={Metal--insulator transition in a weakly interacting many-electron
  system with localized single-particle states},
        date={2006},
     journal={Ann. Physics},
      volume={321},
       pages={1126\ndash 1205},
}


\bib{BK05}{article}{
      author={Bourgain, J.},
      author={Kenig, C.E.},
       title={On localization in the continuous Anderson-Bernoulli model in higher dimension},
        date={2005},
     journal={Invent. Math.},
      volume={161},
       pages={389\ndash 426},
}

\bib{BCSS10}{article}{
author={Boutet de Monvel, A.},
      author={Chulaevsky, V.},
      author={Stollmann, P.},
      author={Suhov, Y.},
       title={Wegnwer-type bounds for a multi-particle continuous Anderson model
with an alloy-type external potential},
        date={2010},
     journal={Journ. Stat. Physics},
      volume={138},
       pages={553\ndash 566},
}

\bib{C10a}{article}{
      author={Chulaevsk{y}, V.},
       title={A remark on charge transfer processes in multi-particle systems},
        date={2010},
     journal={Preprint, \texttt{arXiv:math-ph/1005.3387}},
}


\bib{C12a}{article}{
      author={Chulaevsky, V.},
       title={Direct scaling analysis of localization in single-particle
  quantum systems on graphs with diagonal disorder},
        date={2012},
     journal={Math. Phys. Anal. Geom.},
      volume={15},
       pages={361\ndash 399},
}

\bib{C12b}{article}{
      author={Chula{e}vsky, V.},
       title={On resonances in disordered multi-particle systems},
        date={2011},
     journal={C. R. Acad. Sci. Paris, Ser. I },
      volume={350},
       pages={81\ndash 85},
}

\bib{C12c}{article}{
      author={Chulaevsk{y}, V.},
       title={ Direct Scaling Analysis of localization in disordered systems. II. Multi-particle lattice systems},
        date={2012},
     journal={Preprint, \texttt{arXiv:math-ph/1106.2234}},
}


\bib{C13a}{article}{
      author={Chulaev{s}k{y}, V.},
       title={On the regularity of the conditional distribution of the sample mean},
        date={2013},
     journal={Preprint, \texttt{arXiv:math-ph/1304.6913}},
}

\bib{C14}{article}{
      author={Chulaevsky, V.},
       title={From fixed-energy localization analysis to dynamical localization: An elementary path},
        date={2014},
     journal={J. Stat. Phys.},
      volume={154},
       pages={1391\ndash 1429},
}

\bib{CBS11}{article}{
      author={Chulaevsky, V.},
      author={Boutet~de Monvel, A.},
      author={Suhov, Y.},
       title={Dynamical localization for a multi-particle model with an
  alloy-type external random potential},
        date={2011},
     journal={Nonlinearity},
      volume={24},
       pages={1451\ndash 1472},
}

\bib{CS08}{article}{
      author={Chulaevsky, V.},
      author={Suhov, Y.},
       title={Wegner bounds for a two-particle tight binding model},
        date={2008},
     journal={Commun. Math. Phys.},
      volume={283},
       pages={479\ndash 489},
}

\bib{CS09a}{article}{
      author={Chulaevsky, V.},
      author={S{u}hov, Y.},
       title={Eigenfunctions in a two-particle Anderson tight binding model},
        date={2009},
     journal={Commun. Math. Phys.},
      volume={289},
       pages={701\ndash 723},
}

\bib{CS09b}{article}{
      author={Chulaevsky, V.},
      author={Suhov, Y.},
       title={Multi-particle {A}nderson {l}ocalisation: {i}nduction on the {n}umber of {p}articles},
        date={2009},
     journal={Math. Phys. Anal. Geom.},
      volume={12},
       pages={117\ndash 139},
}

\bib{CS13}{book}{
      author={Chulaevsky, V.},
      author={Su{h}ov, Y.},
       title={Multi-scale Analysis for Random Quantum Systems with
  Interaction},
      series={Progress in Mathematical Physics}
   publisher={Boston: Birkh\"auser},
        date={2013},
}



\bib{DK89}{article}{
      author={Dreifus, H.~von},
      author={Klein, A.},
       title={A new proof of localization in the {A}nderson tight
  binding model},
        date={1989},
     journal={Commun. Math. Phys.},
      volume={124},
      number={7},
       pages={285\ndash 299},
}



\bib{ETV10}{article}{
      author={Elgart, A.},
      author={Tautenhahn, M.},
      author={Veseli\'c, I.},
       title={Anderson localization for a class of models with a sign-indefinite single-site potential via fractional moment method},
        date={2010},
     journal={Ann. Henri Poincar\'e},
      volume={12},
      number={8},
       pages={1571--1599},
}




\bib{FW14}{article}{
      author={Fauser, M.},
      author={Warzel, S.},
       title={Multiparticle localization for disordered systems on continuous space via the fractional moment method},
        date={2014},
     journal={Preprint, \texttt{arXiv:math-ph/1304.6913}},
}

\bib{GK01}{article}{
      author={Germinet, F.},
      author={Klein, A.},
       title={Bootstrap multi-scale analysis and localization in random media},
        date={2001},
     journal={Commun. Math. Physics},
      volume={222},
       pages={415\ndash 448},
}




\bib{GorMP05}{article}{
      author={Gornyi, I.V.},
      author={Mirlin, A.D.},
      author={Polyakov, D.G.},
       title={Interacting electrons in disordered wires: Anderson localization
  and low-temperature transport},
        date={2005},
     journal={Phys. Rev. Lett.},
      volume={95},
       pages={206603},
}

\bib{HK13}{article}{
      author={Hislop, P.D.},
      author={Klopp, F.},
       title={Optimal Wegner estimate and the density of states for n-body,
  interacting Schr\"odinger operators with random potentials},
        date={2013},
     journal={Preprint, \texttt{arXiv:math-ph/1310.6959}},
}


\bib{KN13a}{article}{
      author={Klein, A.},
      author={Nguyen, S.~T.},
       title={Bootstrap multiscale analysis for the multi-particle {A}nderson   model},
        date={2013},
     journal={J. Stat. Phys.},
      volume={151},
      number={5},
       pages={938\ndash 973},
}

\bib{KN13b}{article}{
      author={Klein, A.},
      author={Nguye{n}, S.~T.},
       title={Bootstrap multiscale analysis and localization for multi-particle
  continuous {A}nderson {H}amiltonians},
        date={2013},
     journal={Preprint, {\texttt{arXiv:math-ph/1311.4220}},},
}



\bib{MS85}{article}{
      author={Martinelli, F.},
      author={Scoppola, E.},
       title={ Remark on the absence of absolutely continuous spectrum for d-dimensional Schr\"{o}dinger operators with
                random potential for large disorder or low energy},
        date={1985},
     journal={Commun. Math. Phys.},
      volume={97},
       pages={465\ndash 471},
}

\bib{Sab13}{article}{
      author={Sabri, M.},
       title={Anderson localization for a multi-particle quantum graph},
        date={2014},
     journal={Rev. Math. Phys.},
      volume={26},
      number={1},
      doi={10.1142/S0129055X13500207},
      pages={},
}

\end{biblist}
\end{bibdiv}
\end{document}